\documentclass{article}
\usepackage{amsmath}
\usepackage{amsthm}
\usepackage[margin=1in]{geometry}
\usepackage{graphicx}
\usepackage{authblk}
\usepackage[colorlinks,bookmarksopen,bookmarksnumbered,citecolor=blue,urlcolor=blue,linkcolor=blue]{hyperref}
\usepackage{natbib}
\usepackage{booktabs}
\usepackage[normalem]{ulem}
\usepackage[dvipsnames]{xcolor}
\usepackage{multirow}
\usepackage{pdflscape}
\usepackage{threeparttablex}
\usepackage{longtable}

\usepackage{placeins}

\title{Bayesian joint modelling using semiparametric accelerated failure time approaches}
\author[1]{Ding Ma\thanks{ding.ma@uq.edu.au}}
\author[1]{Patrick Maher}
\author[1,2,3]{Andrew Martin}
\affil[1]{The University of Queensland's Clinical Trials Capability (ULTRA), Brisbane, Australia}
\affil[2]{NHMRC Clinical Trial Centre, University of Sydney, Sydney, Australia}
\affil[3]{Australian and New Zealand Urogenital and Prostate (ANZUP) Cancer Trials Group, Sydney, Australia}

\begin{document}

\maketitle

\begin{abstract}
In longitudinal clinical studies, repeated measurements of biomarkers or health-related quality of life are often collected together with a time-to-event outcome. These two processes are intrinsically linked: longitudinal trajectories may be predictive of event risk, while the occurrence of the event, or its anticipation, can induce informative censoring of the longitudinal process. Joint models provide a principled framework for handling this dependence, but most existing formulations rely on proportional hazards assumptions that may be restrictive and offer limited interpretability on the time scale. We propose a class of semiparametric accelerated failure time joint models that directly model covariate effects on event timing while flexibly capturing longitudinal–event associations. The survival component is specified through an accelerated failure time model with the baseline component represented using a flexible basis expansion, allowing the incorporation of a broad class of smooth baseline specifications. To illustrate this general framework, we consider baseline representations based on Bernstein polynomials, and introduce rescaling strategies to improve numerical stability and parameter identifiability in the presence of time-warping. Estimation is conducted within a Bayesian framework, enabling joint inference for longitudinal, survival, and association parameters. Simulation studies reflecting realistic longitudinal trajectories, censoring mechanisms, and dependence structures are used to evaluate finite-sample performance. The proposed joint models demonstrate improved recovery of longitudinal treatment effects compared with a standalone linear mixed model when event risk depends on the underlying longitudinal process.
Overall, the proposed framework extends existing joint modelling methodology by offering a flexible and interpretable alternative to proportional hazards–based approaches.
\end{abstract}


\newpage

\section{Introduction}

Clinical trials and cohort studies frequently collect longitudinal measurements of biomarkers or health-related quality of life (HRQoL) together with a time-to-event (TTE) outcome. These processes are intrinsically linked: longitudinal trajectories may inform event risk, while event occurrence or anticipation can curtail follow-up, inducing informative censoring. Such dependence is common in oncology and other chronic disease settings, where deterioration in biomarkers or patient-reported outcomes may precede clinical progression. 

Accounting for this interdependence between longitudinal and event processes can be important for valid inference. Joint models address this challenge by linking a mixed-effects submodel for longitudinal trajectories with a survival submodel for the TTE outcome. This framework accounts for measurement error and informative censoring in the longitudinal outcome while providing estimates of how the longitudinal trajectory relates to event timing \citep{wulfsohn1997joint, henderson2000joint, tsiatis2004joint, rizopoulos2012joint}. 

Although joint models are well established for incorporating longitudinal biomarkers as endogenous time-varying covariates in TTE analyses, they are less frequently applied to longitudinal HRQoL data, where linear mixed models (LMMs) remain popular. LMMs assume that data are missing at random, an assumption that could be violated when HRQoL assessments cease at progression or when symptoms worsen \citep{rubin1987statistical}. Using simulation studies, \cite{touraine2023joint} demonstrated that under informative dropout mechanisms, joint models can mitigate bias in longitudinal treatment effect estimates compared to LMMs.

The survival component most often specified with joint models relies on the proportional hazards (PH) model or, less commonly, the proportional odds (PO) model. These formulations underpin most existing joint modelling software, such as \texttt{JM} \citep{rizopoulos2010jm}, \texttt{JMbayes} \citep{rizopoulos2016r}, \texttt{JMbayes2} \citep{rizopoulos2025jmbayes2}, and \texttt{JSM} \citep{xu2020semi}. Outside of a joint framework, these formulations assume baseline covariate effects that are constant over time—PH on the hazard scale and PO on the log–cumulative-odds scale. Within a joint model, however, event risk depends not only on baseline covariates but also on the evolving longitudinal process; consequently, apparent departures from proportionality in marginal hazard patterns may arise even when proportionality holds conditionally.

Alongside PH and PO formulations, accelerated failure time (AFT) models offer a complementary way to characterise event-time data by expressing covariate effects directly on the time scale. By quantifying how covariates accelerate or decelerate the expected time to an event, AFT models provide time-based summaries that may align well with clinical interpretation of outcomes such as time to progression or duration of response. Traditional parametric AFT models specify a particular distribution for event times (e.g., Weibull, log-normal, or log-logistic) to define the baseline survival, density and hazard functions. While convenient, these assumptions can be overly restrictive when the true event-time distribution is complex or unknown, potentially leading to model misspecification and bias. To overcome this limitation, semiparametric AFT (sAFT) models extend the framework by retaining the regression structure linking covariates and longitudinal processes to event time while estimating the baseline TTE function nonparametrically.

sAFT formulations have been explored in survival analysis via a range of estimation approaches, including rank-based methods \citep{prentice1978linear, jin2003rank, chiou2014fitting}, least-squares-based methods \citep{buckley1979linear, jin2006least} and full-likelihood-based methods (e.g., \cite{komarek2005accelerated, ZenLin07, crowther2022flexible, ma2023semiparametric}). However, only a limited number of studies have extended the sAFT model to the joint modelling framework, with all of these employing a full likelihood form (for likelihood-based inference).
Moreover, each of these studies specifies the sAFT structure through the hazard function. \cite{TeHsWa05} approximate the baseline hazard in the sAFT submodel using piecewise-constant functions, which results in a spiky likelihood and substantial computational challenges. \cite{tseng2016kernel} mitigate these issues by introducing a kernel-smoothed baseline hazard.
To our knowledge, existing sAFT joint models have largely focused on likelihood-based, hazard-driven formulations. This common choice reflects the substantial technical challenges involved in embedding longitudinal processes within log-linear AFT representations, particularly when flexible baseline specifications are desired.

In this article, we develop a semiparametric accelerated failure time joint model that allows flexible, modular baseline representations. The remainder of the article is organised as follows.
Section~\ref{sec:jm-structure} introduces the joint model structure. Section~\ref{sec:bayes} describes the Bayesian estimation approach.
Section~\ref{sec:application} applies the method to HRQoL and progression data from ENZAMET, a large randomised trial in metastatic hormone-sensitive prostate cancer \citep{sweeney2023testosterone}, using these data solely to illustrate the methodology rather than for inferential purposes; this example also motivates the simulation study in Section~\ref{sec:simulation}. Finally, Section~\ref{sec:conclusion} discusses these results and gives concluding remarks.

\section{Joint model structure} \label{sec:jm-structure}

We adopt a shared‐parameter framework in which the longitudinal and TTE processes are linked through subject-specific random effects \citep{rizopoulos2012joint}, implying that the two processes are conditionally independent given these effects. We first specify the longitudinal submodel that describes repeated measurements over time, followed by the survival submodel that governs event occurrence. The two components are then connected through a functional association term, yielding a joint likelihood for inference.

\subsection{Longitudinal submodel}

We begin by introducing the longitudinal submodel used to describe each individual’s repeated measurements. Let individual $i$ be observed at visit times $0 = t_{i0} < t_{i1} < \cdots < t_{in_i} = t_i$, where $t_i$ denotes the follow-up time. Longitudinal measurements are collected intermittently, and the record from the final time point $t_i$ does not contribute to estimation because the process is taken to be right-continuous.

Let $\mathbf{x}_i(t)$ denote the $1 \times p$ vector of fixed-effect covariates for individual $i$ at time $t$. The corresponding observed covariate vector at the $j$th visit is $\mathbf{x}_{ij} = \mathbf{x}_i(t_{ij})$. Collecting the $n_i$ observed vectors for individual $i$ yields the $n_i \times p$ matrix $\mathbf{X}_i$. Stacking observations from all individuals produces the $N \times p$ matrix $\mathbf{X}$, where $N = \sum_{i=1}^n n_i$. The associated fixed-effect coefficients are represented by the $p \times 1$ vector $\boldsymbol{\beta}$.

Similarly, let $\mathbf{z}_i(t)$ denote the $1 \times q$ vector of covariates corresponding to the random effects, with $\mathbf{z}_{ij} = \mathbf{z}_i(t_{ij})$ being the value observed at the $j$th visit. Each individual has an associated $q \times 1$ random-effects vector $\boldsymbol{b}_i$, assuming $\boldsymbol{b}_i \sim \mathcal{N}(\mathbf{0},\boldsymbol{\Sigma_b})$, $\forall i = 1,2,\dots,n$. The residual error process is denoted $\varepsilon_i(t)$, with $\varepsilon_i(t) \sim \mathcal{N}(0, \sigma_{\varepsilon}^2)$, and $\varepsilon_{ij} = \varepsilon_i(t_{ij})$ as its discrete-time counterpart. 

In this article, we adopt a simple specification for the longitudinal process and model it using an LMM formulation, specifically,
\begin{equation} \label{eq: longitudinal submodel}
     y_i(t) = y_i^*(t) + \varepsilon_i(t) = \mathbf{x}_i(t) \boldsymbol{\beta} + \mathbf{z}_i(t) \boldsymbol{b}_i + \varepsilon_i(t) ,
\end{equation}
with the corresponding discrete representation given by
\begin{equation*}
    y_{ij} = y_{ij}^* + \varepsilon_{ij} = \mathbf{x}_{ij} \boldsymbol{\beta} + \mathbf{z}_{ij} \boldsymbol{b}_i + \varepsilon_{ij} .
\end{equation*}
With this defined, an individual longitudinal likelihood is represented via the product of normally distributed individual outcomes
\begin{equation}
\label{eq:longitudinal-likelihood}
p\bigl(\mathbf{y}_i \mid \boldsymbol{\beta}, \boldsymbol{b}_i, \sigma_{\varepsilon}^2, \mathbf{X}_i, \mathbf{Z}_i\bigr)
    = \prod_{j=1}^{n_i}
      \mathcal{N}\!\left(
        y_{ij} \mid
        \mathbf{x}_{ij} \boldsymbol{\beta}
        + \mathbf{z}_{ij} \boldsymbol{b}_i,
        \ \sigma_{\varepsilon}^2
      \right),
\end{equation}
with $\boldsymbol{y}_i = (y_{i1},\dots,y_{in_i})^\top$ as the longitudinal measurements across $n_i$ time points for individual $i$.

\subsection{Survival submodel}

We now introduce the AFT model for the TTE observations.
The classical AFT model, expressed through hazard functions, is typically written as
\begin{equation*} \label{eq: classical aft}
    \lambda_i(t) = \lambda_0(e^{- \mathbf{w}_i \boldsymbol{\gamma}} t) e^{- \mathbf{w}_i \boldsymbol{\gamma}} ,
\end{equation*}
with $\mathbf{w}_i$ being the (time-invariant) covariate vector contributing to the survival process (e.g., treatment allocation, age), and $\boldsymbol{\gamma}$ as the corresponding coefficient vector.
It is important to emphasise that incorporating time-varying covariate effects into AFT models, whether arising directly from exogenous covariates or indirectly through endogenous longitudinal processes, results in substantially greater complexity in both model specification and estimation (as discussed in later sections), exceeding that encountered in Cox PH models. In this context, the accelerated failure time $\kappa_i(t)$ depends on the entire history of the underlying longitudinal trajectory of $y_i^*$ up to time $t$, denoted by $\widetilde{\mathbf{y}}_i^*(t) = \{y_i^*(s): 0 \leq s < t\}$.
The contribution of the longitudinal submodel to the survival submodel is incorporated through a transformation $g$, yielding an extension to the classical expression,
\begin{equation} \label{eq: sAFT-JM}
    \lambda_i(t \mid g) = \lambda_0(\kappa_i(t)) e^{- \mathbf{w}_i \boldsymbol{\gamma} - \alpha g}
\end{equation}
with 
\begin{equation} \label{eq: accelerated failure time with g}
    \kappa_i(t) = \int_0^t e^{- \mathbf{w}_i \boldsymbol{\gamma} - \alpha g} ds .
\end{equation}
$g$ may be defined broadly: it can be time-varying, reflecting longitudinal process; or time-invariant, for example when based on random effects that do not vary over time. Typically, in a longitudinal setting as considered in this article, $g$ is a function of time.

For the functional form linking the two submodels, we adopt the ``current value'' specification \citep{rizopoulos2012joint}, i.e. $g = y_i^*(t)$. Accordingly, when incorporating the time-varying covariate effects from the longitudinal submodel, equations \eqref{eq: sAFT-JM} and \eqref{eq: accelerated failure time with g} are modified to
\begin{equation*} \label{eq: longitudinal aft}
    \lambda_i(t \mid \widetilde{\mathbf{y}}_i^*(t)) = \lambda_0(\kappa_i(t)) e^{- \mathbf{w}_i \boldsymbol{\gamma} - \alpha y_i^*(t)}
\end{equation*}
and
\begin{equation} \label{eq: accelerated failure time with longitudinal effect}
    \kappa_i(t) = \int_0^t e^{- \mathbf{w}_i \boldsymbol{\gamma} - \alpha y_i^*(s)} ds .
\end{equation}

The survival likelihood for individual $i$ is
\begin{equation} \label{eq: aft likelihood using baseline hazard}
    p(t_i, \delta_i \mid \boldsymbol{\gamma}, \alpha, \lambda_0, \mathbf{w}_i, \widetilde{\mathbf{y}}_i^*(t))
    = {\lambda_i(t_i)}^{\delta_i} S_i(t_i)
    = \left\{ \lambda_0(\kappa_i(t_i)) \exp(- \mathbf{w}_i \boldsymbol{\gamma} - \alpha y_i^*(t_i)) \right\}^{\delta_i} \exp \left( -\Lambda_0(\kappa_i(t_i)) \right) ,
\end{equation}
where the cumulative baseline hazard is denoted by $\Lambda_0(\kappa_i(t)) = \int_{0}^{\kappa_i(t)} \lambda_0(s) ds$.

The baseline hazard $\lambda_0$ is assumed to be a smooth function over time, although typically only a finite number of observations are available in real scenarios.
Following the theoretical framework of Grenander's method of sieves \citep{Grenander81, GeHw82} and the implementation described in \cite{ma2023semiparametric}, we consider approximating $\lambda_0$ using a linear combination of basis functions,
\begin{equation} \label{eq: baseline hazard approx}
    \lambda_0(\kappa_i(t)) = \sum_{u=1}^{m} \theta_u \phi_u(\kappa_i(t)) ,
\end{equation}
with the corresponding cumulative baseline hazard given by
\begin{equation} \label{eq: cumulative baseline hazard approx}
    \Lambda_0(\kappa_i(t)) = \sum_{u=1}^{m} \theta_u \Phi_u(\kappa_i(t)),
\end{equation}
where the integrated basis function is defined by $\Phi_u(\kappa_i(t)) = \int_{0}^{\kappa_i(t)} \phi_u(s) ds$.

However, in an AFT model, the baseline hazard is a function of $\kappa_i(t)$, rather than a function of $t$ directly as in a Cox PH model.
The expression in \eqref{eq: accelerated failure time with longitudinal effect} indicates that the values of $\kappa_i(t)$ depend on the time-invariant coefficients $\boldsymbol{\gamma}$, the association parameter $\alpha$ and the underlying longitudinal trajectory $\widetilde{\mathbf{y}}_i^*(t)$.
During computation, the estimates of $\boldsymbol{\gamma}$, $\alpha$ and $\widetilde{\mathbf{y}}_i^*(t)$ are updated iteratively, which in turn alters both the realised values and the overall range of $\kappa_i(t)$. Such variation can impede stable convergence of the sampler.

To mitigate this issue, we adopt the rescaling strategy in \cite{panaro2020spsurv}. Specifically, in each iteration, we apply an additional normalisation step to $\kappa_i(t)$.
Defining $\kappa_i^* = \frac{\kappa_i(t)}{M}$ with $M = \max \{\kappa_i(t)\}$ in this iteration, so that $\kappa_i^* \in (0, 1]$ before approximating the baseline and cumulative baseline hazards. Under this transformation, with some simple derivation, equations \eqref{eq: baseline hazard approx} and \eqref{eq: cumulative baseline hazard approx} become
\begin{equation} \label{eq: baseline hazard approxs rescaled}
    \lambda_0(\kappa_i(t)) = \dfrac{1}{M} \sum_{u=1}^{m} \theta_u \phi_u(\kappa_i^*) \quad \text{and} \quad \Lambda_0(\kappa_i(t)) = \sum_{u=1}^{m} \theta_u \Phi_u(\kappa_i^*) .
\end{equation}

The non-negativity of the baseline hazard, together with the non-negativity and monotonicity of the cumulative baseline hazard, is guaranteed by employing basis functions with non-negative support and by imposing priors on the coefficients $\theta_u$ whose support is restricted to the nonnegative real line. Optional regularisation may be applied to $\theta_u$ to promote extra smoothness in the approximated baseline hazard functions. This article considers Bernstein polynomials (as adopted in \cite{panaro2020spsurv}) as the bases,
\begin{equation*}
    \phi_u(\kappa_i^*) = \binom{m-1}{u-1} (\kappa_i^*)^{u-1} ( 1 - \kappa_i^*)^{m - u}
    \quad \text{and} \quad
    \Phi_u(\kappa_i^*) = \frac{1}{m} \sum_{v=u}^{m} \binom{m}{v} (\kappa_i^*)^v (1-\kappa_i^*)^{m-v},
\end{equation*}
which are defined over the unit interval and require only the specification of degrees of freedom $m$. The selection of $m$ can be guided by the number of events $e$ motivated by the method of sieves approximation for $m = \lceil e^{1/3}\rceil$ \citep{ghosal2001convergence}. This ensures that the complexity of the Bernstein polynomial approximation increases as the amount of event information increases. From this, sensitivity analyses can be conducted as exploration of neighbouring $m$ values for fine tuning. Misspecification of $m$ can bias and oversmooth in cases of too few bases, with overfitting and instability arising for too many degrees of freedom, highlighting the need for careful selection of $m$.

Substituting \eqref{eq: baseline hazard approxs rescaled} into \eqref{eq: aft likelihood using baseline hazard}, the survival likelihood of individual $i$ becomes
\begin{equation} \label{eq: survival-likelihood}
p(t_i, \delta_i \mid \boldsymbol{\gamma}, \alpha, \boldsymbol{\theta}, \mathbf{w}_i, \widetilde{\mathbf{y}}_i^*(t))
= \left\{ \left[ \dfrac{1}{M} \sum_{u=1}^{m} \theta_u \phi_u(\kappa_i^*) \right] \exp(- \mathbf{w}_i \boldsymbol{\gamma} - \alpha y_i^*(t_i)) \right\}^{\delta_i} \exp \left( - \sum_{u=1}^{m} \theta_u \Phi_u(\kappa_i^*) \right) .
\end{equation}
The longitudinal likelihood \eqref{eq:longitudinal-likelihood} and the survival likelihood \eqref{eq: survival-likelihood} are combined within the Bayesian inference framework described in the next section.

\section{Bayesian estimation} \label{sec:bayes}

\subsection{Likelihood and priors}

Inference for the semi-parametric joint model is conducted within a fully Bayesian framework. Let
$$\boldsymbol{\Theta}
=
\left(
  \boldsymbol{\beta},
  \boldsymbol{\gamma},
  \alpha,
  \boldsymbol{\theta},
  \{\boldsymbol{b}_i\}_{i=1}^n,
  \boldsymbol{L_{\Sigma_b}},
  \sigma_\varepsilon^2
\right)$$
denote the complete collection of parameters. With $\boldsymbol{\theta}$ as the basis weights for the baseline hazard approximation, $\boldsymbol{\gamma}$ as survival fixed effects, $\alpha$ as the association parameter linking the longitudinal and survival submodels, $\boldsymbol{\beta}$ denoting longitudinal fixed effects, $\{\boldsymbol{b}_i\}_{i=1}^n$ as the random effects vectors for the $n$ participants, $\boldsymbol{L_{\Sigma_b}}$ as the Cholesky factor of the random effects covariance matrix $\boldsymbol{\Sigma_{b}}$ for efficient decomposition, and finally $\sigma_\varepsilon^{2}$ as the residual variance.

Following from assumptions of independence for both the longitudinal \eqref{eq: longitudinal submodel} and survival sub-models \eqref{eq: survival-likelihood}, the full joint likelihood together with the prior distributions defines the joint posterior distribution:
\begin{equation} \label{eq:full-posterior}
\begin{aligned}
p(\boldsymbol{\Theta}\mid \mathcal{D}) \propto & \left[\prod_{i=1}^n p(\mathbf{y}_i \mid \boldsymbol{\beta},\boldsymbol{b}_i,\sigma_\varepsilon^2,\mathbf{X}_i,\mathbf{Z}_i)\;
p(t_i, \delta_i \mid \boldsymbol{\gamma}, \alpha, \boldsymbol{\theta}, \mathbf{w}_i, \widetilde{\mathbf{y}}_i^*(t))\; p(\boldsymbol{b}_i \mid \boldsymbol{L_{\Sigma_b}}) \right]\; \\ 
& p(\boldsymbol{\beta})\,p(\boldsymbol{\gamma})\,p(\alpha)\, p(\boldsymbol{\theta})\, p(\boldsymbol{L_{\Sigma_b}})\,p(\sigma_\varepsilon^2).
\end{aligned}
\end{equation}
This joint framework joins information via shared random effects; the factorisation explicitly shows such dependence. By integrating out random effects, one obtains the marginal posterior distribution:
\begin{equation*}
p(\boldsymbol{\beta},\boldsymbol{\gamma},\alpha,\boldsymbol{\theta},\boldsymbol{L_{\Sigma_b}},\sigma_\varepsilon^2 \mid \mathcal{D}) \propto \prod_{i=1}^n\int p(\mathbf{y}_i \mid \boldsymbol{b}_i,\cdot)\, p(t_i,\delta_i \mid \boldsymbol{b}_i,\cdot)\, p(\boldsymbol{b}_i\mid \boldsymbol{L_{\Sigma_b}})\,d\boldsymbol{b}_i \times \text{(priors)}.
\end{equation*}

Weakly informative priors were used, allowing data to dominate posterior inference whilst maintaining computational stability and ensuring proper posterior distributions. Regression coefficients in both the longitudinal and survival submodels were assigned normal priors centred at zero with large variances.

The basis weights $\boldsymbol{\theta}$ were constrained to be non-negative, thereby ensuring positivity while remaining weakly informative in scale. The random-effects covariance matrix $\boldsymbol{\Sigma}_b$ was parameterised via a Cholesky decomposition, which facilitates an efficient separation of marginal variances and correlation structure. Priors were specified to induce moderate correlations among the random effects, reflecting a priori plausible dependence in HRQoL measurements across visits within individuals. The residual variance was also assigned a weakly informative prior, calibrated to preserve a well-behaved posterior distribution. 

The association parameter $\alpha$ is the key parameter linking the longitudinal and survival submodels. We assigned $\alpha$ a weakly informative normal prior centred at zero, with the scale chosen to stabilise estimation without imposing strong prior information. Prior sensitivity analyses were conducted to assess the impact of this scale; the results, presented in the Supplementary Material, showed little change in posterior estimation across the range of prior scales considered.

For assistance in modelling decisions of likelihood specification and prior tuning, guidance on model selection and diagnostic procedures are detailed within the Supplementary Material.

\subsection{Posterior computation}

As is commonly seen within Bayesian survival analysis, the joint posterior distribution in \eqref{eq:full-posterior} does not admit a closed-form solution. Thereby, MCMC sampling was performed using the No-U-Turn Sampler (NUTS) \citep{hoffman2014nuts}, a form of Hamiltonian Monte Carlo implemented in \texttt{Stan} \citep{carpenter2017stan}. NUTS acted as an appropriate sampling method for the complex and high-dimensional parameter spaces found within these joint modelling frameworks. 

We employed 1,000 warm-up iterations and 1,000 sampling iterations across four MCMC chains. Convergence was assessed via trace plots, $\hat{R}$ diagnostics, and effective sample sizes (ESS) \citep{vehtari_rhat_ess}. Specifically, convergence was concluded when trace plots appeared stable, all parameters had $\hat{R} < 1.01$, and ESS exceeded 400 for both bulk and tail quantities.

Several parameterisation strategies were implemented to enhance computational stability and efficiency. Random effects $\mathbf{b}_i$ were parameterised using a non-centred formulation, with $\mathbf{b}_i = \boldsymbol{L_{\Sigma_b}} \mathbf{z}_{b,i}$, where $\mathbf{z}_{b,i} \sim \mathcal{N}(\mathbf{0}, \mathbf{I}_2)$ and $\boldsymbol{L_{\Sigma_b}}$ is the Cholesky factor of the covariance matrix $\boldsymbol{\Sigma}_b$ such that $\boldsymbol{\Sigma}_b = \boldsymbol{L_{\Sigma_b}} \boldsymbol{L_{\Sigma_b}}^\top$.
This decomposition reduces correlation between hyperparameters and random effects, thereby improving the efficiency of random-effect sampling. Likelihood contributions for both longitudinal and survival components were computed on the log scale, enhancing numerical stability by summing log-likelihood increments rather than multiplying probabilities.

\section{ENZAMET example} \label{sec:application}

ENZAMET randomised 1{,}125 men with metastatic hormone-sensitive prostate cancer (mHSPC) to receive testosterone suppression plus enzalutamide or testosterone suppression plus a standard nonsteroidal antiandrogen \citep{davis2019enzalutamide, stockler2022health, sweeney2023testosterone}. The primary endpoint of overall survival in ENZAMET showed no important departures from the proportional hazards assumption. For the secondary endpoint of clinical progression-free survival (cPFS), different diagnostic approaches gave somewhat mixed indications. Visual checks of scaled Schoenfeld residuals showed a reasonably stable pattern over time, whereas the Grambsch–Therneau test \citep{grambsch1994proportional} ($p=0.013$) and the score-process approach of Lin, Wei, and Ying \citep{lin1993checking} ($p=0.025$) suggested possible non-proportional hazards.  

We used data from the subset of 1,105 ENZAMET participants who provided HRQoL information to demonstrate our proposed sAFT-JM and not to re-evaluate the trial’s established findings. HRQoL was measured at baseline and at regular follow-up clinic visits until disease progression using multiple instruments and scales, but we focus solely on the Global Health Status/Quality of Life scale (which provides an overall summary score ranging from 0 to 100). Because HRQoL assessments ceased at the time of clinical progression, the availability of longitudinal measurements is inherently linked to the event process, motivating the use of JM in this setting.

For the longitudinal submodel, we employed the LMM structure described in Section~\ref{sec:jm-structure}, specifying fixed effects for time and an arm-by-time interaction, together with subject-specific random intercepts and slopes as in \cite{touraine2023joint}:

\begin{equation} \label{eq: ENZAMET longitudinal model} 
    y_i(t) = y_i^*(t) + \varepsilon_i(t) = \beta_0 + \beta_1 t + \beta_2 \cdot 
    (\text{arm}_i \cdot t) + b_{0i} + b_{1i} t + \varepsilon_i(t) .
\end{equation}
The survival submodel incorporated treatment effects, with the longitudinal process entering through the current value of the underlying HRQoL trajectory:
\begin{equation} \label{eq: longitudinal aft applied}
    \lambda_i(t \mid \widetilde{\mathbf{y}}_i^*(t)) = \lambda_0(\kappa_i(t)) e^{- \text{arm}_i \cdot \gamma - \alpha y_i^*(t)} .
\end{equation}

We adopted the Bayesian approach from Section~\ref{sec:bayes} to fit the sAFT-JM, using Bernstein polynomial bases with five degrees of freedom. Table \ref{tab:ENZAMET-results} summarises the longitudinal and joint model estimates of health-related quality of life (HRQoL) over follow-up. 

A small but consistent decline in HRQoL over time was observed, with the monthly slope parameter $\beta_{1}$ estimated at $-0.038$ under the sAFT-JM and $-0.039$ under the LMM. The interaction between treatment and time ($\beta_{2}$) was negative under both models, indicating a slightly steeper decline in HRQoL in the treatment group. Estimates of the random intercept standard deviation ($\sigma_{b_0} \approx 14.5$), random slope standard deviation ($\sigma_{b_1} \approx 0.20$), and residual variability ($\sigma_{\varepsilon} \approx 11.6$) were nearly identical across modelling frameworks.

In the sAFT-JM, the estimated association parameter $\alpha$ provided evidence of a positive association between the longitudinal HRQoL process and survival time, while the direct treatment effect on the survival scale, $\gamma$, was also estimated to be positive. 

\begin{table}[ht]
\centering
\begin{threeparttable}
\caption{ENZAMET Results}
\label{tab:ENZAMET-results}

\begin{tabular}{lcc}
\toprule
\textbf{} 
  & \textbf{sAFT-JM}
  & \textbf{LMM} \\
\midrule

$\beta_{0}$  
& 72.921 (72.017, 73.865)   
& 72.922 (72.014, 73.830)  \\[4pt]

$\beta_{1}$  
& -0.038 (-0.069, -0.007)  
& -0.039 (-0.069, -0.009)  \\[4pt]

$\beta_{2}$  
& -0.042 (-0.079, -0.005)   
& -0.038 (-0.075, -0.000)  \\[4pt]

$\sigma_{b_0}$  
& 14.549 (13.868, 15.236)  
& 14.528 (13.859, 15.230)  \\[4pt]

$\sigma_{b_1}$  
& 0.205 (0.185, 0.226)  
& 0.204 (0.184, 0.225) \\[4pt]

$\sigma_{\varepsilon}$  
& 11.581 (11.450, 11.712)  
& 11.581 (11.450, 11.712)  \\[6pt]

$\gamma$  
&  0.898 (0.732, 1.068)  
&  \\[4pt]

$\alpha$  
& 0.011 (0.005, 0.017)  
&  \\

\bottomrule
\end{tabular}

\begin{tablenotes}
\small
\item Posterior mean with 95\% credible intervals for sAFT-JM, mean with 95\% confidence intervals for LMM. Empty cells reflect parameters not estimated under the LMM.
\end{tablenotes}

\end{threeparttable}
\end{table}

\section{Simulation} \label{sec:simulation}

We followed the simulation study framework of \cite{morris2019using} to evaluate and compare the operating characteristics of the proposed sAFT-JM and a standalone LMM. The data-generating mechanisms, including the assumed longitudinal and TTE processes and their association, are described in Section~\ref{sec:Data Generation Mechanisms}. The primary targets of the simulation study were the longitudinal fixed effects $\beta_{0}, \beta_{1}, \beta_{2}$ and the TTE and association parameters $\gamma$ and $\alpha$. Model performance was evaluated in terms of bias, root mean squared error, and credible interval coverage across repeated simulations.

\subsection{Data generation mechanism} \label{sec:Data Generation Mechanisms}

We used the functional form of the longitudinal submodel in \eqref{eq: ENZAMET longitudinal model}. The true HRQoL score trajectory, $\widetilde{\mathbf{y}}_i^*(t)$, is assumed to follow a linear function of time, with each observation given by:
\begin{equation} \label{eq: touraine mixed model}
    y_i^*(t) = \beta_0 + \beta_1 t + \beta_2 \cdot 
    (\text{arm}_i \cdot t) + b_{0i} + b_{1i} t ,
\end{equation}
where $\beta_0$, $\beta_1$ and $\beta_2$ are the coefficients of fixed effects; $b_{0i}$ and $b_{1i}$ are the coefficients of random effects, $\left(\begin{smallmatrix}
  b_{0i} \\
  b_{1i}
\end{smallmatrix}\right) \sim \mathcal{N} \left( \mathbf{0}, \left(\begin{smallmatrix}
  \sigma_0^2 & \sigma_{01} \\
  \sigma_{01} & \sigma_1^2
\end{smallmatrix}\right) \right)$.
Given the fixed effect coefficients $\beta_0$, $\beta_1$ and $\beta_2$, the random effect coefficients $b_{0i}$ and $b_{1i}$, and sequence of visit times $\{t_{ij}\}$ for each individual ($\{t_{ij}\}$ is allowed to vary in both values and length across individuals), we can generate the trajectory of observed HRQoL score, $\widetilde{\mathbf{y}}_i(t)$. 

We adopted the same survival submodel as \eqref{eq: longitudinal aft applied}, considering the treatment effect and the true HRQoL score trajectory (via ``current value'' link), which is
\begin{equation*}
    \lambda_i(t) = \lambda_0(\kappa_i(t)) e^{- \text{arm}_i \cdot \gamma - \alpha y_i^*(t)}
\end{equation*}
with
\begin{equation} \label{eq: accelerated failure time simulation}
    \kappa_i(t) =
    \begin{cases}
        e^{-C_1} \cdot \dfrac{1 - e^{-C_2 t}}{C_2} & \text{if} \ \ C_2 \neq 0;
        \\
        e^{-C_1} t & \text{if} \ \ C_2 = 0 ,
    \end{cases}
\end{equation}
where we denote $C_1 := \text{arm}_i \cdot \gamma + \alpha(\beta_0 + b_{0i})$ and $C_2 := \alpha (\beta_1 + \beta_2 \cdot \text{arm}_i + b_{1i})$. It is important to note that \eqref{eq: accelerated failure time simulation} is derived under the assumption that the exact functional form of the true HRQoL score trajectory specified in \eqref{eq: touraine mixed model} is known during data generation.

The way to generate event time is intuitive, simply calculating the inverse $t = S^{-1}(\xi)$ with $\xi \sim \text{Unif}(0,1)$. We selected the log-logistic distribution to represent unimodal hazard shape, as inspired by the ENZAMET example in the previous section. The baseline survival function assuming a log-logistic distribution is
\begin{equation} \label{eq: survival function loglogistic}
    S_i(t) = S_0(\kappa_i(t)) = \dfrac{1}{1 + \left( \dfrac{\kappa_i(t)}{\varsigma}\right)^a} ,
\end{equation}
with $a$ being the shape parameter and $\varsigma$ being the scale parameter.

Plug \eqref{eq: accelerated failure time simulation} into \eqref{eq: survival function loglogistic} and carry the inverse calculation, we can obtain the closed-form solution for event time
\begin{equation*}
\label{eq: loglogistic_event_time}
    t_i =
    \begin{cases}
        - \dfrac{1}{C_2} \log \left( 1 - C_2 \cdot \varsigma \cdot e^{C_1} \left( \frac{1}{\xi} - 1 \right)^{1/a} \right) & \text{if} \ \ C_2 \neq 0; \\
        \varsigma \cdot e^{C_1} \left( \frac{1}{\xi} - 1 \right)^{1/a} & \text{if} \ \ C_2 = 0 .
    \end{cases}
\end{equation*}
In addition, we considered the Weibull distribution to represent monotonic hazard shape, where the baseline survival function is
\begin{equation} \label{eq: survival function weibull}
    S_i(t) = S_0(\kappa_i(t)) = \exp \left(- \left( \dfrac{\kappa_i(t)}{\varsigma}\right)^a \right),
\end{equation}
with $a$ and $\varsigma$ being the shape parameter and scale parameter, respectively. $0 < a < 1$ represents decreasing hazard over time, $a = 1$ represents constant hazard (Weibull distribution reduces to exponential distribution), and $a > 1$ represents increasing hazard. We focus on the decreasing and increasing hazard situations in this article.
The relevant Weibull event time is therefore,
\begin{equation}
    t_i =
    \begin{cases}
        - \dfrac{1}{C_2} \log \left( 1 - \varsigma \cdot C_2 \cdot e^{C_1} \left( -\log \xi \right)^{1/a} \right) & \text{if} \ \ C_2 \neq 0; \\
        \varsigma \cdot e^{C_1} \left( -\log \xi \right)^{1/a} & \text{if} \ \ C_2 = 0 .
    \end{cases}
\end{equation}

\subsubsection{Censoring} \label{sec:censoring}

Two censoring mechanisms are considered in this article. Under the first mechanism, denoted as $\text{CM}_1$, an administrative censoring time $c_{\text{adm}}$ is imposed, yielding the observed follow-up time $t_i^{*} = \min(t_i, c_{\text{adm}})$ and censoring indicator $\delta_i = \mathbf{1}_{\{t_i \leq c_{\text{adm}}\}}$.
Under the second mechanism, denoted as $\text{CM}_2$, we impose both an administrative censoring time $c_{\text{adm}}$ and an additional random censoring time $c_i$ with $c_i$ following an exponential distribution. The observed follow-up time then becomes $t_i^{*} = \min(t_i, c_i, c_{\text{adm}})$ with censoring indicator $\delta_i = \mathbf{1}_{\{t_i \leq \min(c_i, c_{\text{adm}})\}}$. The event time $t_i$, administrative censoring time $c_{\text{adm}}$, and random censoring time $c_i$ are mutually independent.

\subsubsection{Simulation inputs} \label{sec:sim inputs}

Simulation settings were chosen to broadly align with the ENZAMET example. A total sample size of $n=1{,}100$ patients was assumed throughout. 
The HRQoL assessment schedule was represented on a month-based scale, with baseline, month~1, and then 3-monthly assessments thereafter, corresponding to the original 4-week and 12-weekly assessment schedule under a 4-weeks-per-month approximation. To reflect variability in assessment times, scheduled visits were jittered. For patient~$i$, independent Gaussian jitter with mean 0 and standard deviation 1 month was added to each scheduled visit time. Baseline assessments were fixed at time 0, and negative jittered times were truncated to 0.

Longitudinal HRQoL outcomes were generated according to
\eqref{eq: ENZAMET longitudinal model}.
Survival times were generated under four parametric AFT baseline distributions chosen to reflect a range of baseline hazard shapes motivated by the fitted sAFT-JM. The primary log-logistic setting used shape parameter $a = 1.20$ and scale parameter $\varsigma = 23$. Additional Weibull settings used scale parameter $38$ with shape parameters $0.90$, $1.30$, and $2.10$, representing decreasing, concave increasing, and convex increasing baseline hazard shapes, respectively. Longitudinal HRQoL measurements were retained only if their assessment time occurred before the observed follow-up time $t_i^{*}$.

The coefficient scenarios considered are summarised in Table~\ref{tab:simulation_scenarios_params}. Scenarios differed only in the assumed treatment effects on the longitudinal HRQoL trajectory and the TTE outcome; all other model parameters were held fixed and based on rounded Bernstein sAFT-JM estimates from Table~\ref{tab:ENZAMET-results}. Together, the scenarios were designed to span null, concordant, and discordant treatment-effect regimes.

\begin{table}[ht]
\centering
\begin{threeparttable}
\caption{Coefficient scenarios used in simulations}
\label{tab:simulation_scenarios_params}

\begin{tabular}{c c c c c c c c}
\hline
\multirow{2}{*}{\textbf{Scenario\tnote{a}}} 
& \multicolumn{2}{c}{\textbf{Effect of Treatment On:}}
& \multirow{2}{*}{$\boldsymbol{\beta_{1}}$\tnote{b}}
& \multirow{2}{*}{$\boldsymbol{\beta_{2}}$}
& \multirow{2}{*}{$\boldsymbol{\beta_{1} + \beta_{2}}$\tnote{c}}
& \multirow{2}{*}{$\boldsymbol{\gamma}$}
& \multirow{2}{*}{$\boldsymbol{\exp(\gamma)}$\tnote{d}} \\
\cline{2-3}
& \textbf{TTE}
& \textbf{HRQoL decline}
& 
& 
& 
& 
& \\
\hline
1 & -- & -- & $-0.04$ & $0.00$  & $-0.04$ & $0.00$  & $1.00$ \\
\hline
2 & $\uparrow$ & $\uparrow$ & $-0.04$ & $0.04$  & $0.00$  & $0.90$  & $2.46$ \\
3 & $\uparrow$ & $\downarrow$ & $-0.04$ & $-0.04$ & $-0.08$ & $0.90$  & $2.46$ \\
\hline
4 & $\downarrow$ & $\uparrow$ & $-0.04$ & $0.04$  & $0.00$  & $-0.90$ & $0.41$ \\
5 & $\downarrow$ & $\downarrow$ & $-0.04$ & $-0.04$ & $-0.08$ & $-0.90$ & $0.41$ \\

\hline
\end{tabular}

\begin{tablenotes}
\small
\item[a] All scenarios used $\alpha=0.012$, $\beta_0=73$, $\sigma_{b_0}=15$, $\sigma_{b_1}=0.2$, and $\sigma_\varepsilon=12$, based on rounded Bernstein sAFT-JM estimates (Table~\ref{tab:ENZAMET-results}). Arrows indicate the qualitative direction of the treatment effect relative to control: $\uparrow$ improvement, $\downarrow$ worsening, and -- no change. Scenarios were designed to represent null, concordant, and discordant regimes of treatment effects on HRQoL and TTE. 
\item[b] $\beta_1$ denotes the monthly rate of HRQoL change in the control arm.
\item[c] $\beta_1 + \beta_2$ denotes the total monthly rate of HRQoL change in the treatment arm.
\item[d] $\exp(\gamma)$ represents the acceleration factor for cPFS comparing treatment to control; values greater than~1 indicate prolonged cPFS and values less than~1 indicate shortened cPFS.
\end{tablenotes}

\end{threeparttable}
\end{table}

Two censoring mechanisms, described in Section~\ref{sec:censoring}, were applied to each scenario. Administrative censoring was imposed at month~120, which yielded censoring proportions below 50\% across all five scenarios and therefore allowed a separate setting with approximately 50\% overall censoring. Specifically, independent exponential censoring times were generated as $c_i \sim \mathrm{Exp}(\lambda_c)$, with the rate parameter $\lambda_c$ calibrated to achieve approximately 50\% overall censoring.

In total, 40 data-generation configurations were used to evaluate the proposed sAFT-JM and compare its performance with that of the standalone LMM. These configurations were formed by crossing four event-time distributions, five coefficient scenarios, and two censoring mechanisms, with 1,000 replicates generated for each configuration.

\subsection{Prior specification and initial values}

For the proposed sAFT-JM, the longitudinal response was centred by subtracting the observed mean HRQoL value in each simulated dataset before model fitting. This centring was used only to improve numerical stability; posterior summaries for the longitudinal intercept were transformed back to the original HRQoL scale. The fixed-effect priors for both the longitudinal and survival regression coefficients were centred at zero, with that the preliminary fits were used only to construct initial values.

Initial values for the longitudinal parameters were obtained from a preliminary linear mixed model fitted using \texttt{nlme::lme()}, with fixed effects for time and the treatment-by-time interaction, together with patient-specific random intercepts and time slopes. Initial values for the survival treatment coefficient were obtained from an exponential AFT model fitted using \texttt{survival::survreg()} with \texttt{dist = "exponential"}. Small independent jitter was added to these preliminary estimates across chains. The random-effect correlation matrix was initialised with correlation 0.2. The association parameter was initialised near zero. The Bernstein polynomial baseline weights were initialised as small positive values, drawn from a normal distribution with mean 0.1 and standard deviation 0.01, truncated to be positive.

The prior distributions used in the proposed sAFT-JM are summarised in Table~\ref{tab:simulation_priors}. The association parameter was assigned a weakly informative normal prior, $\alpha \sim N\{0, [\log(2)/1.96]^2\}$. This prior corresponds approximately to allowing a two-fold multiplicative change in event time over the calibration contrast. For numerical stability, this prior was implemented using an equivalent scaled parameterisation in Stan. Sensitivity to the prior scale for $\alpha$ was examined in an additional simulation analysis, with detailed results reported in the Supplementary
Material.

\begin{table}[!ht]
\centering
\caption{Prior distributions used for the proposed sAFT-JM in the simulation study.}
\label{tab:simulation_priors}
\begin{tabular}{lll}
\hline
Parameter & Description & Prior distribution \\
\hline
$\beta_0, \beta_1, \beta_2$ &
Longitudinal fixed effects &
$N(0, 10^2)$ \\

$\gamma$ &
AFT treatment coefficient in the survival submodel &
$N(0, 10^2)$ \\

$\alpha$ &
Association parameter &
$N\{0, [\log(2)/1.96]^2\}$ \\

$\theta_1,\ldots,\theta_5$ &
Bernstein polynomial baseline weights &
$N(0, 5^2)$, truncated below at 0 \\

$\sigma_e$ &
Longitudinal residual standard deviation &
Half-Cauchy$(0, 5)$ \\

$\sigma_{b0}, \sigma_{b1}$ &
Random intercept and random slope standard deviations &
Half-Cauchy$(0, 5)$ \\

$\Omega_b$ &
Random-effect correlation matrix &
LKJ correlation prior with shape 2 \\
\hline
\end{tabular}
\end{table}

In applications to new data, we recommend using weakly informative priors centred at clinically neutral values, with scales chosen on the scale of the corresponding parameter. For longitudinal fixed effects, prior scales should reflect the measurement scale of the outcome and the expected magnitude of change over follow-up. For AFT regression coefficients, prior scales should be interpreted on the time-ratio scale, since $\exp(\gamma)$ represents the multiplicative effect on event time. For the association parameter, priors should be calibrated to plausible changes in event time over clinically meaningful contrasts in the longitudinal outcome. When limited prior information is available, sensitivity analyses over narrower and wider prior scales are recommended, particularly for the association parameter and baseline hazard parameters.

\subsection{Simulation results}

\begin{landscape}
\begin{figure}[p]
\centering
\includegraphics[width=\linewidth, height=\textheight, keepaspectratio]{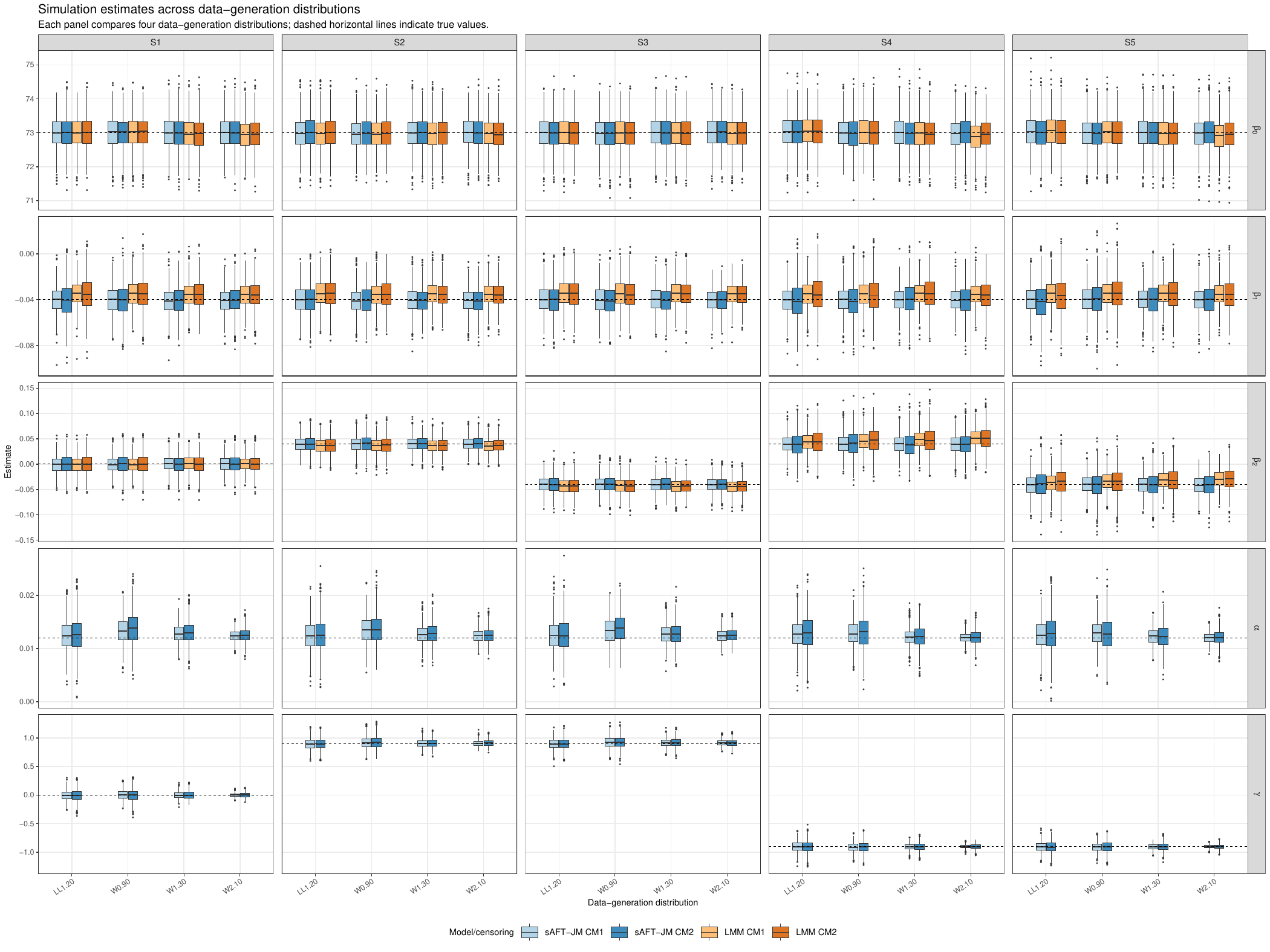}
\caption{Simulation results for Scenarios 1 to 5. Box plots show per-simulation estimates from sAFT-JM (blue boxes) and LMM (orange boxes) across baseline specifications and censoring mechanisms.}
\label{fig:scenario_coefficient_boxplot_matrix_loglogistic}
\end{figure}
\end{landscape}

Table~\ref{tab:saft-lmm-distribution-estimates} summarises the mean estimates across the 40 data-generation configurations, and Figure~\ref{fig:scenario_coefficient_boxplot_matrix_loglogistic} shows the corresponding boxplots of estimates across simulation replicates. Detailed bias, RMSE, empirical standard deviation, posterior standard deviation, and coverage summaries are provided in the Supplementary Material. Overall, the proposed sAFT-JM recovered the longitudinal fixed effects well across the four event-time distributions, five coefficient scenarios, and two censoring mechanisms. The longitudinal intercept $\beta_0$ was estimated accurately by both the proposed joint model and the standalone LMM, with only small deviations from the true value of 73 across all settings.

The main difference between the methods was observed for the longitudinal time slope $\beta_1$. Across scenarios and baseline event-time distributions, the standalone LMM tended to estimate $\beta_1$ closer to zero than the true value of $-0.04$, reflecting attenuation of the average HRQoL decline when informative event-related dropout was ignored. In contrast, the sAFT-JM produced estimates of $\beta_1$ that were consistently close to the data-generating value under both administrative censoring and controlled 50\% censoring.

For the treatment-by-time interaction $\beta_2$, the sAFT-JM also tracked the data-generating values closely across the five coefficient scenarios. The standalone LMM showed more scenario-dependent bias, with the largest deviations observed in Scenarios 4 and 5, where the treatment effect on event time was negative. In these settings, shortened event times led to greater truncation of the longitudinal process, particularly among treated patients, amplifying the informative-dropout mechanism. This pattern is consistent with the expected consequence of analysing the longitudinal measurements alone: when event risk depends on the underlying HRQoL trajectory, a standalone LMM can distort estimates of treatment-related longitudinal change. 

The survival-side coefficients were estimated only under the proposed sAFT-JM. The AFT treatment coefficient $\gamma$ was generally recovered well across baseline event-time distributions. Estimates were close to zero in the null scenario and close to the true values of $0.9$ and $-0.9$ in the remaining coefficient scenarios. The association parameter $\alpha$ was also estimated stably, with mean estimates close to the data-generating value of $0.012$. A mild tendency toward overestimation was observed, but the mean estimates remained close to the truth across the simulation settings. Small deviations were observed across some baseline distributions and censoring mechanisms, but the overall pattern in Table~\ref{tab:saft-lmm-distribution-estimates} and Figure~\ref{fig:scenario_coefficient_boxplot_matrix_loglogistic} indicates that the fitted joint model captured the longitudinal--event association reasonably well.

The comparison between administrative censoring and controlled 50\% censoring showed that heavier censoring increased variability in some settings but did not materially change the qualitative conclusions. Across the simulation study, the proposed sAFT-JM provided improved recovery of longitudinal treatment-effect parameters relative to the standalone LMM, while also estimating the survival treatment effect and longitudinal--event association within the same modelling framework.

We also examined computational burden for the standalone LMM, the Cox-based joint model, and the proposed sAFT-JM. All simulation computations were performed on Bunya, The University of Queensland's high-performance computing cluster, using EPYC3 and EPYC4 CPU nodes. Across the 40 simulation configurations, the standalone LMM was computationally negligible, with mean, median, and 95th percentile runtimes of 0.005, 0.004, and 0.008 minutes per simulated dataset, respectively. The corresponding runtimes for the Cox-JM fitted using \texttt{JMbayes2} were 1.2, 1.1, and 1.7 minutes, while those for the proposed sAFT-JM were 27.2, 21.4, and 67.2 minutes. These results indicate that the proposed model is more computationally demanding. However, this comparison should be interpreted in light of both implementation and model complexity: \texttt{JMbayes2} provides an optimised implementation of established joint models, whereas the proposed sAFT-JM was fitted using a custom Stan implementation and involves additional computational challenges, including repeated updating of the accelerated time scale during posterior sampling.

Two supplementary analyses were conducted to assess robustness of these conclusions. First, under the log-logistic data-generation settings, longitudinal estimates from the standalone LMM, the proposed sAFT-JM, and a Cox-JM were compared; both joint modelling approaches reduced the attenuation in longitudinal slope estimates observed under the standalone LMM. Second, sensitivity analyses for the prior scale of $\alpha$ showed little material change in the estimated longitudinal, survival, or association parameters. Detailed results for both analyses are provided in the Supplementary Material.

\begin{landscape}
\begin{table}[!htbp]
\centering
\footnotesize
\caption{sAFT-JM and LMM estimates by data-generation distribution.}
\label{tab:saft-lmm-distribution-estimates}
\begin{tabular}{llrrrrrrrrrrrrrrrrr}
\toprule
Scen. & Par. & True & \multicolumn{4}{c}{LL1.20} & \multicolumn{4}{c}{W0.90} & \multicolumn{4}{c}{W1.30} & \multicolumn{4}{c}{W2.10}
\\
\cmidrule(lr){4-7} \cmidrule(lr){8-11} \cmidrule(lr){12-15} \cmidrule(lr){16-19}
 &  &  & \multicolumn{2}{c}{$\text{CM}_1$} & \multicolumn{2}{c}{$\text{CM}_2$} & \multicolumn{2}{c}{$\text{CM}_1$} & \multicolumn{2}{c}{$\text{CM}_2$} & \multicolumn{2}{c}{$\text{CM}_1$} & \multicolumn{2}{c}{$\text{CM}_2$} & \multicolumn{2}{c}{$\text{CM}_1$} & \multicolumn{2}{c}{$\text{CM}_2$}
\\
\cmidrule(lr){4-5} \cmidrule(lr){6-7} \cmidrule(lr){8-9} \cmidrule(lr){10-11} \cmidrule(lr){12-13} \cmidrule(lr){14-15} \cmidrule(lr){16-17} \cmidrule(lr){18-19}
 &  &  & sAFT & LMM & sAFT & LMM & sAFT & LMM & sAFT & LMM & sAFT & LMM & sAFT & LMM & sAFT & LMM & sAFT & LMM
\\
\midrule
1 & $\beta_0$ & 73.000 & 72.998 & 73.000 & 73.007 & 73.011 & 73.003 & 73.014 & 73.004 & 73.021 & 73.011 & 72.983 & 72.985 & 72.967 & 73.009 & 72.949 & 73.002 & 72.960 \\
 & $\beta_1$ & -0.040 & -0.040 & -0.035 & -0.040 & -0.035 & -0.040 & -0.035 & -0.041 & -0.035 & -0.041 & -0.035 & -0.041 & -0.035 & -0.041 & -0.036 & -0.041 & -0.036 \\
 & $\beta_2$ & 0.000 & -0.001 & -0.001 & 0.000 & 0.000 & -0.001 & -0.001 & 0.000 & 0.000 & 0.001 & 0.001 & -0.000 & -0.000 & 0.001 & 0.001 & 0.000 & 0.000 \\
 & $\alpha$ & 0.012 & 0.012 & -- & 0.013 & -- & 0.013 & -- & 0.014 & -- & 0.013 & -- & 0.013 & -- & 0.012 & -- & 0.013 & -- \\
 & $\gamma$ & 0.000 & -0.004 & -- & -0.003 & -- & -0.000 & -- & -0.001 & -- & -0.001 & -- & -0.003 & -- & -0.000 & -- & -0.001 & -- \\
\addlinespace
2 & $\beta_0$ & 73.000 & 72.990 & 72.986 & 73.018 & 73.016 & 72.975 & 72.982 & 72.992 & 73.001 & 72.993 & 72.971 & 73.021 & 73.002 & 73.014 & 72.975 & 73.006 & 72.968 \\
 & $\beta_1$ & -0.040 & -0.040 & -0.035 & -0.040 & -0.034 & -0.041 & -0.035 & -0.041 & -0.035 & -0.041 & -0.035 & -0.041 & -0.035 & -0.041 & -0.035 & -0.041 & -0.036 \\
 & $\beta_2$ & 0.040 & 0.040 & 0.037 & 0.040 & 0.037 & 0.040 & 0.037 & 0.041 & 0.038 & 0.040 & 0.037 & 0.041 & 0.037 & 0.040 & 0.036 & 0.041 & 0.037 \\
 & $\alpha$ & 0.012 & 0.012 & -- & 0.013 & -- & 0.013 & -- & 0.014 & -- & 0.013 & -- & 0.013 & -- & 0.012 & -- & 0.013 & -- \\
 & $\gamma$ & 0.900 & 0.892 & -- & 0.898 & -- & 0.917 & -- & 0.921 & -- & 0.904 & -- & 0.909 & -- & 0.908 & -- & 0.912 & -- \\
\addlinespace
3 & $\beta_0$ & 73.000 & 73.001 & 72.999 & 72.989 & 72.990 & 72.984 & 72.995 & 72.970 & 72.983 & 73.010 & 72.993 & 72.996 & 72.982 & 73.011 & 72.976 & 73.018 & 72.985 \\
 & $\beta_1$ & -0.040 & -0.040 & -0.034 & -0.040 & -0.035 & -0.040 & -0.035 & -0.041 & -0.036 & -0.040 & -0.034 & -0.041 & -0.035 & -0.040 & -0.035 & -0.040 & -0.035 \\
 & $\beta_2$ & -0.040 & -0.040 & -0.043 & -0.040 & -0.043 & -0.040 & -0.043 & -0.040 & -0.042 & -0.041 & -0.045 & -0.039 & -0.043 & -0.040 & -0.044 & -0.040 & -0.044 \\
 & $\alpha$ & 0.012 & 0.013 & -- & 0.013 & -- & 0.013 & -- & 0.014 & -- & 0.013 & -- & 0.013 & -- & 0.012 & -- & 0.013 & -- \\
 & $\gamma$ & 0.900 & 0.896 & -- & 0.898 & -- & 0.927 & -- & 0.927 & -- & 0.914 & -- & 0.918 & -- & 0.913 & -- & 0.912 & -- \\
\addlinespace
4 & $\beta_0$ & 73.000 & 73.028 & 73.044 & 73.024 & 73.039 & 73.008 & 73.030 & 72.981 & 73.008 & 73.007 & 72.980 & 72.975 & 72.961 & 72.965 & 72.890 & 73.006 & 72.956 \\
 & $\beta_1$ & -0.040 & -0.040 & -0.035 & -0.041 & -0.036 & -0.040 & -0.035 & -0.041 & -0.036 & -0.040 & -0.035 & -0.039 & -0.035 & -0.040 & -0.035 & -0.040 & -0.036 \\
 & $\beta_2$ & 0.040 & 0.040 & 0.044 & 0.039 & 0.044 & 0.040 & 0.046 & 0.041 & 0.047 & 0.040 & 0.049 & 0.039 & 0.047 & 0.040 & 0.052 & 0.040 & 0.051 \\
 & $\alpha$ & 0.012 & 0.013 & -- & 0.013 & -- & 0.013 & -- & 0.013 & -- & 0.012 & -- & 0.012 & -- & 0.012 & -- & 0.012 & -- \\
 & $\gamma$ & -0.900 & -0.904 & -- & -0.908 & -- & -0.909 & -- & -0.914 & -- & -0.906 & -- & -0.907 & -- & -0.903 & -- & -0.905 & -- \\
\addlinespace
5 & $\beta_0$ & 73.000 & 73.042 & 73.060 & 73.005 & 73.022 & 72.992 & 73.016 & 72.977 & 73.005 & 73.004 & 72.977 & 72.987 & 72.975 & 72.982 & 72.908 & 73.022 & 72.971 \\
 & $\beta_1$ & -0.040 & -0.040 & -0.034 & -0.042 & -0.036 & -0.040 & -0.035 & -0.039 & -0.035 & -0.040 & -0.034 & -0.040 & -0.035 & -0.040 & -0.035 & -0.040 & -0.036 \\
 & $\beta_2$ & -0.040 & -0.041 & -0.037 & -0.039 & -0.034 & -0.040 & -0.034 & -0.041 & -0.035 & -0.041 & -0.032 & -0.041 & -0.032 & -0.042 & -0.029 & -0.041 & -0.029 \\
 & $\alpha$ & 0.012 & 0.013 & -- & 0.013 & -- & 0.013 & -- & 0.013 & -- & 0.012 & -- & 0.012 & -- & 0.012 & -- & 0.012 & -- \\
 & $\gamma$ & -0.900 & -0.906 & -- & -0.911 & -- & -0.910 & -- & -0.909 & -- & -0.907 & -- & -0.907 & -- & -0.902 & -- & -0.905 & -- \\
\bottomrule
\end{tabular}%
\vspace{0.5ex}
\begin{minipage}{0.95\linewidth}
\footnotesize Notes: $\text{CM}_1$ = administrative censoring; $\text{CM}_2$ = controlled 50\% censoring. LL1.20 = log-logistic(1.20, 23.0); W0.90, W1.30, and W2.10 = Weibull(shape, 38.0). sAFT denotes the proposed sAFT-JM. Entries are means of point estimates across simulations; dashes indicate parameters not fitted by the standalone LMM.
\end{minipage}
\end{table}

\end{landscape}

\FloatBarrier

\section{Discussion and conclusion} \label{sec:conclusion}

We developed a sAFT-JM for the joint analysis of longitudinal and time-to-event (TTE) data, demonstrated its application in an analysis of HRQoL and cPFS from the ENZAMET trial, and evaluated its performance through a comprehensive simulation study. The main contribution is the integration of a semiparametric AFT joint modelling formulation with a Bayesian estimation framework and a flexible basis-expanded baseline event-time component, allowing posterior inference for longitudinal, survival, association, and baseline parameters within a single model.

The results presented here should be interpreted as characterising the operating behaviour of an early-phase implementation of the proposed semiparametric AFT joint modelling framework, rather than as providing definitive performance guarantees across all settings. As with many flexible joint modelling approaches, numerical behaviour and inferential properties depend on modelling choices for the baseline event-time component, the strength of longitudinal–event association, and tuning parameters used in estimation. The focus of this work is therefore on establishing the feasibility of the proposed framework in realistic clinical settings, while transparently documenting its operating characteristics.

Consistent with previous findings \citep{touraine2023joint}, the simulation results highlight that ignoring the dependence between longitudinal outcomes and event risk can bias longitudinal treatment effect estimates. Across scenarios, longitudinal fixed effects were recovered accurately under the joint model, even in the presence of informative dropout, reinforcing the value of joint modelling approaches in settings where longitudinal measurements are truncated by informative event processes.

The ENZAMET illustration demonstrates how joint models can be used in settings where HRQoL assessments cease at clinical progression, leading to informative dropout that may affect standard longitudinal analyses. In such scenarios, jointly modelling HRQoL and cPFS provides a principled framework for accounting for this dependence when estimating longitudinal trajectories. This illustration does not reflect on the ENZAMET HRQoL findings themselves, which used deterioration-free survival as the prespecified primary endpoint and are therefore not subject to this issue \citep{stockler2022health}. The same joint-modelling framework, of which the proposed semiparametric AFT formulation is one variant, can also be applied to biomarkers, such as prostate-specific antigen in prostate cancer. In that context, joint models can help distinguish treatment effects mediated through biomarker dynamics from those acting independently, providing a useful tool for investigating treatment mechanisms, evaluating surrogacy, and improving dynamic prediction.

A key strength of the proposed approach is the use of an accelerated failure time formulation, which yields treatment effects directly on the time scale and may be more interpretable than relative hazards in some clinical settings. The use of a semiparametric baseline hazard allows flexibility without imposing strong parametric assumptions, and the simulation design incorporated realistic longitudinal variability and censoring patterns.

This work also has limitations. First, only a single association structure (current value) was examined, and alternative association structures may exhibit different operating characteristics. Second, the longitudinal submodel was specified as a Gaussian linear mixed model, which may be restrictive for bounded HRQoL outcomes when skewness, heteroscedasticity, floor or ceiling effects, or boundary inflation are present.
Third, although the proposed sAFT-JM generally recovered both longitudinal and survival-side parameters well across the simulation settings, the association parameter showed a mild tendency toward overestimation. This bias was small relative to the data-generating value and did not materially affect the main longitudinal treatment-effect conclusions, but it suggests that absolute estimates of the longitudinal--event association should be interpreted with appropriate caution.
Finally, the proposed model incurred greater computational cost than the standalone LMM and Cox-JM comparator, reflecting both the mathematical complexity of the semiparametric AFT joint model and the current custom implementation.

Future work should consider alternative association structures, extensions of the longitudinal component beyond the Gaussian linear mixed model, and further refinement of the baseline hazard approximation, including scaling, regularisation, and sensitivity to the choice of basis family. The supplementary Cox-JM analysis was intended as a targeted comparison of longitudinal-side estimation under an alternative joint-modelling framework, rather than as a direct comparison of survival-side estimands, since AFT and proportional hazards models parameterise covariate effects on different scales. Further work could extend this idea by comparing longitudinal-side inference across joint-modelling frameworks under carefully defined data-generating mechanisms, while avoiding direct interpretation of AFT time-ratio effects and Cox hazard-ratio effects as equivalent quantities. Future work should also examine subject-specific dynamic prediction under the proposed sAFT-JM, including assessment of calibration, discrimination, and prediction error. In parallel, computational optimisation and more efficient posterior sampling strategies will be important for improving practical usability. A longer-term aim is to develop an R package that provides a more accessible and reproducible implementation of the proposed modelling framework.

In conclusion, the proposed sAFT-based joint model offers a flexible and interpretable framework for the joint analysis of longitudinal and time-to-event data. Its semiparametric AFT structure provides an alternative perspective to established proportional hazards–based joint models, particularly in settings where time-scale interpretations are appealing. The approach should be viewed as complementary rather than superior, and ongoing work aims to further clarify its operating characteristics and optimal implementation in practice.

\newpage

\section*{Acknowledgements}
The authors gratefully acknowledge the Australian and New Zealand Urogenital and Prostate (ANZUP) Cancer Trials Group and the NHMRC Clinical Trials Centre at the University of Sydney for access to the dataset used in this study. Ding Ma’s work was partially supported by funding from the Australian Trials Methodology (AusTriM) Research Network, a National Health and Medical Research Council (NHMRC) Centre of Research Excellence (Grant ID 1171422).

\section*{Data availability statement}

A reproducible example is publicly accessible on GitHub via \url{https://github.com/UQ-ULTRA/sAFT-JM-BiomJ}.
The ENZAMET clinical trial data were not generated by the authors and cannot be shared directly. Researchers seeking access to the ENZAMET data should submit their request to the ENZAMET Trial Executive, with applications made via the Australia and New Zealand Urogenital and Prostate (ANZUP) Cancer Trials Group, which coordinates data access and governance. \\

\section*{Conflict of interest}
The authors declare no potential conflicts of interest.

\bibliographystyle{apalike}
\bibliography{references.bib}

\end{document}


\maketitle

\noindent
This Supplementary Material provides additional details supporting the simulation study reported in the main manuscript. It includes model-checking and information-criterion details, detailed operating-characteristic summaries for the main 40-configuration simulation, a targeted log-logistic comparison of longitudinal estimates across the LMM, sAFT-JM, and Cox-JM, and sensitivity analyses for the prior scale of the association parameter \(\alpha\).

\section{Model comparison and calibration}
\label{sec:supp-model-checking}

In this section, we outline guidance for prior selection and methods for model calibration through information criteria and diagnostics, with elaboration on approaches readers may consider when choosing across models.

\subsection{Predictive checks}
\label{sec:supp-predictive-checks}

Prior predictive checks were conducted to assess whether the specified priors yielded plausible longitudinal trajectories and baseline hazard shapes. This was achieved by sampling parameter values $\boldsymbol{\Theta}^{(s)} \sim p(\boldsymbol{\Theta})$ and generating replicated datasets $\mathcal{D}^{\mathrm{rep},(s)}$ for both the longitudinal and survival components. Visual inspection of the resulting HRQoL trajectories and event-time distributions confirmed that the implied data-generating mechanisms lay within realistic ranges, thereby supporting the adequacy of the prior specifications.

Posterior predictive checks were similarly performed to evaluate the agreement between the observed data and data generated from the fitted models. Specifically, for each posterior draw $\boldsymbol{\Theta}^{(s)} \sim p(\boldsymbol{\Theta} \mid \mathcal{D})$, a replicated dataset $\mathcal{D}^{\mathrm{rep}}$ was generated, and relevant summary statistics were compared with those computed from the observed data.

\subsection{Information criteria and comparison}
\label{sec:supp-information-criteria}

For formal model comparison, we recommend multiple information criteria that may be used to jointly assess model fit and complexity. The Deviance Information Criterion (DIC) is a widely used Bayesian measure that balances goodness of fit with an estimate of the effective number of parameters. The deviance used within the DIC is based on the underlying likelihood \citep{Celeux2006}. One information criterion useful within this context is the marginal DIC, obtained via integrating out subject-specific random effects from the full joint likelihood \citep{Celeux2006,Millar2009}, expressed as:
\begin{align*}
D(\boldsymbol{\Psi}) = -2 \sum_{i=1}^{n} \log \left[ \int p(\mathbf{y}_i \mid \boldsymbol{b}_i, \boldsymbol{\Psi})\, p(t_i, \delta_i \mid \boldsymbol{b}_i, \boldsymbol{\Psi})\, p(\boldsymbol{b}_i \mid \boldsymbol{\Psi}) \, d\boldsymbol{b}_i \right].
\end{align*}
This is then used to define the degrees of freedom $p_D$ to construct the DIC with:
\begin{align*}
p_D = \bar{D} - D(\bar{\boldsymbol{\Psi}})
\qquad \text{and} \qquad
\mathrm{DIC} = D(\bar{\boldsymbol{\Psi}}) + 2p_D,
\end{align*}
with $\bar{D} = E\!\left[D(\boldsymbol{\Psi}) \mid \mathcal{D}\right]$ and $\bar{\boldsymbol{\Psi}} = E(\boldsymbol{\Psi}\mid \mathcal{D})$ \citep{spiegelhalter2002dic}. 
We recommend the use of the marginal distribution as our primary objective was population-level inference. However, if one is interested within the hierarchical effects and individual-specific prediction they may consider constructing their information criterion on conditional deviance.  DIC was preferred over the Akaike Information Criterion (AIC) and Bayesian Information Criterion (BIC) because it was specifically developed for complex Bayesian hierarchical models in which the effective number of parameters is not well defined, unlike the fixed penalty structures inherent to AIC and BIC \citep{spiegelhalter2002dic}. Moreover, Spiegelhalter et al. note that AIC may perform poorly under informative priors, as adopted in this joint modelling framework, further motivating the use of DIC.

In addition, the Widely Applicable Information Criterion (WAIC; \citealp{watanabe2010waic}) may be used as an alternative fully Bayesian measure that accounts for the entire posterior distribution rather than relying on point estimates \citep{gelman2014waic}. WAIC is defined as
$\text{WAIC} = -2\left(\text{lppd} - p_{\text{WAIC}}\right)$,
where the log pointwise predictive density is given by $\text{lppd} = \sum_{i=1}^{n} \log \int p(y_i \mid \boldsymbol{\Theta}) p(\boldsymbol{\Theta} \mid \mathcal{D}) \, d\boldsymbol{\Theta}$, and $p_{\text{WAIC}}$ denotes the effective number of parameters estimated from the variance of the log-likelihood across posterior draws.

Alongside DIC and WAIC, approximate leave-one-out cross-validation (LOO-CV) based on Pareto-smoothed importance sampling \citep{vehtari2017loo} offers yet another alternative method to estimate out-of-sample predictive performance. LOO-CV is generally preferred to WAIC in settings with influential observations or substantial posterior uncertainty, as indicated by the Pareto k diagnostics. In particular, a non-negligible proportion of observations with $k > 0.7$ suggests that LOO-CV estimates may be unreliable and that alternative strategies may be required \citep{loo}.

\section{Detailed sAFT-JM Simulation Performance}
\label{sec:supp-saft-performance}

Supplementary Tables~\ref{tab:saft-bp-distribution-details-cm1}--\ref{tab:saft-bp-distribution-details-cm2}
provide detailed operating characteristics for the proposed sAFT-JM under
administrative censoring and controlled 50\% censoring. These summaries support
the conclusions in the main text: longitudinal fixed effects were recovered with
small bias, empirical and posterior standard deviations were generally similar,
and coverage was mostly close to the nominal 95\% level. The association
parameter showed a mild positive bias, but this was small relative to the
data-generating value.

\begin{landscape}
\begin{table}[!htbp]
\centering
\tiny
\caption{Detailed sAFT-JM performance by data-generation distribution under administrative censoring.}
\label{tab:saft-bp-distribution-details-cm1}
\begin{tabular}{llrrrrrrrrrrrrrrrrrrrr}
\toprule
Scen. & Par. & \multicolumn{5}{c}{LL1.20} & \multicolumn{5}{c}{W0.90} & \multicolumn{5}{c}{W1.30} & \multicolumn{5}{c}{W2.10}
\\
\cmidrule(lr){3-7} \cmidrule(lr){8-12} \cmidrule(lr){13-17} \cmidrule(lr){18-22}
 &  & Bias & ESD & PSD & RMSE & CP & Bias & ESD & PSD & RMSE & CP & Bias & ESD & PSD & RMSE & CP & Bias & ESD & PSD & RMSE & CP
\\
\midrule
1 & $\beta_0$ & -0.0022 & 0.4573 & 0.4759 & 0.4571 & 0.969 & 0.0025 & 0.4899 & 0.4761 & 0.4896 & 0.943 & 0.0113 & 0.4761 & 0.4741 & 0.4760 & 0.947 & 0.0093 & 0.4710 & 0.4719 & 0.4709 & 0.952 \\
 & $\beta_1$ & 0.0000 & 0.0114 & 0.0120 & 0.0114 & 0.959 & 0.0000 & 0.0116 & 0.0118 & 0.0116 & 0.952 & -0.0010 & 0.0115 & 0.0113 & 0.0116 & 0.945 & -0.0008 & 0.0108 & 0.0106 & 0.0108 & 0.951 \\
 & $\beta_2$ & -0.0008 & 0.0165 & 0.0166 & 0.0165 & 0.947 & -0.0008 & 0.0159 & 0.0164 & 0.0159 & 0.951 & 0.0010 & 0.0159 & 0.0155 & 0.0159 & 0.941 & 0.0007 & 0.0144 & 0.0146 & 0.0144 & 0.955 \\
 & $\sigma_{b0}$ & 0.0199 & 0.3471 & 0.3580 & 0.3475 & 0.954 & -0.0083 & 0.3877 & 0.3727 & 0.3876 & 0.950 & -0.0249 & 0.4221 & 0.3954 & 0.4226 & 0.965 & -0.0316 & 0.4132 & 0.3891 & 0.4142 & 0.943 \\
 & $\sigma_{b1}$ & 0.0005 & 0.0071 & 0.0070 & 0.0071 & 0.947 & 0.0006 & 0.0080 & 0.0073 & 0.0080 & 0.953 & 0.0006 & 0.0094 & 0.0075 & 0.0094 & 0.937 & 0.0010 & 0.0083 & 0.0071 & 0.0084 & 0.951 \\
 & $\sigma_{\epsilon}$ & -0.0034 & 0.0598 & 0.0608 & 0.0598 & 0.950 & 0.0045 & 0.0758 & 0.0651 & 0.0759 & 0.948 & 0.0071 & 0.0912 & 0.0689 & 0.0914 & 0.944 & 0.0088 & 0.0842 & 0.0670 & 0.0846 & 0.954 \\
 & $\alpha$ & 0.0005 & 0.0028 & 0.0027 & 0.0028 & 0.933 & 0.0013 & 0.0025 & 0.0025 & 0.0028 & 0.915 & 0.0008 & 0.0017 & 0.0017 & 0.0019 & 0.927 & 0.0004 & 0.0010 & 0.0011 & 0.0011 & 0.942 \\
 & $\gamma$ & -0.0041 & 0.0931 & 0.0888 & 0.0932 & 0.941 & -0.0004 & 0.0854 & 0.0810 & 0.0854 & 0.936 & -0.0009 & 0.0553 & 0.0546 & 0.0553 & 0.949 & -0.0002 & 0.0332 & 0.0325 & 0.0332 & 0.942 \\
\addlinespace
2 & $\beta_0$ & -0.0095 & 0.4783 & 0.4726 & 0.4781 & 0.945 & -0.0247 & 0.4618 & 0.4724 & 0.4623 & 0.953 & -0.0075 & 0.4658 & 0.4708 & 0.4657 & 0.956 & 0.0142 & 0.4557 & 0.4694 & 0.4557 & 0.960 \\
 & $\beta_1$ & -0.0000 & 0.0121 & 0.0120 & 0.0121 & 0.952 & -0.0007 & 0.0115 & 0.0118 & 0.0115 & 0.948 & -0.0006 & 0.0112 & 0.0112 & 0.0112 & 0.945 & -0.0007 & 0.0104 & 0.0107 & 0.0105 & 0.949 \\
 & $\beta_2$ & -0.0003 & 0.0153 & 0.0156 & 0.0153 & 0.954 & -0.0000 & 0.0154 & 0.0154 & 0.0154 & 0.946 & 0.0002 & 0.0143 & 0.0146 & 0.0143 & 0.949 & 0.0002 & 0.0136 & 0.0140 & 0.0136 & 0.949 \\
 & $\sigma_{b0}$ & -0.0247 & 0.4512 & 0.3959 & 0.4517 & 0.950 & -0.0298 & 0.4071 & 0.3814 & 0.4080 & 0.948 & -0.0494 & 0.5067 & 0.4267 & 0.5089 & 0.941 & -0.0862 & 0.5578 & 0.4677 & 0.5641 & 0.955 \\
 & $\sigma_{b1}$ & 0.0011 & 0.0082 & 0.0071 & 0.0083 & 0.952 & 0.0004 & 0.0072 & 0.0068 & 0.0072 & 0.959 & 0.0014 & 0.0087 & 0.0073 & 0.0088 & 0.947 & 0.0016 & 0.0086 & 0.0074 & 0.0087 & 0.959 \\
 & $\sigma_{\epsilon}$ & 0.0067 & 0.0857 & 0.0674 & 0.0859 & 0.961 & 0.0051 & 0.0804 & 0.0646 & 0.0805 & 0.934 & 0.0169 & 0.1003 & 0.0740 & 0.1016 & 0.955 & 0.0191 & 0.1171 & 0.0820 & 0.1186 & 0.950 \\
 & $\alpha$ & 0.0004 & 0.0028 & 0.0027 & 0.0029 & 0.933 & 0.0015 & 0.0027 & 0.0026 & 0.0031 & 0.905 & 0.0007 & 0.0018 & 0.0019 & 0.0019 & 0.938 & 0.0005 & 0.0013 & 0.0013 & 0.0013 & 0.937 \\
 & $\gamma$ & -0.0075 & 0.0918 & 0.0932 & 0.0921 & 0.954 & 0.0168 & 0.1018 & 0.0996 & 0.1032 & 0.936 & 0.0036 & 0.0718 & 0.0714 & 0.0718 & 0.949 & 0.0078 & 0.0534 & 0.0536 & 0.0540 & 0.954 \\
\addlinespace
3 & $\beta_0$ & 0.0007 & 0.4684 & 0.4735 & 0.4681 & 0.950 & -0.0158 & 0.4567 & 0.4734 & 0.4568 & 0.954 & 0.0103 & 0.4750 & 0.4708 & 0.4749 & 0.948 & 0.0106 & 0.4788 & 0.4688 & 0.4787 & 0.938 \\
 & $\beta_1$ & 0.0002 & 0.0120 & 0.0120 & 0.0120 & 0.947 & -0.0003 & 0.0122 & 0.0118 & 0.0122 & 0.936 & 0.0003 & 0.0114 & 0.0112 & 0.0114 & 0.952 & -0.0005 & 0.0103 & 0.0106 & 0.0103 & 0.954 \\
 & $\beta_2$ & -0.0005 & 0.0155 & 0.0156 & 0.0155 & 0.952 & 0.0003 & 0.0158 & 0.0154 & 0.0157 & 0.943 & -0.0010 & 0.0149 & 0.0147 & 0.0149 & 0.947 & 0.0001 & 0.0137 & 0.0139 & 0.0137 & 0.945 \\
 & $\sigma_{b0}$ & -0.0162 & 0.4678 & 0.4044 & 0.4678 & 0.943 & -0.0127 & 0.3788 & 0.3753 & 0.3789 & 0.960 & -0.0133 & 0.4344 & 0.3991 & 0.4344 & 0.956 & -0.0689 & 0.5175 & 0.4498 & 0.5219 & 0.960 \\
 & $\sigma_{b1}$ & 0.0010 & 0.0083 & 0.0072 & 0.0083 & 0.945 & 0.0007 & 0.0076 & 0.0068 & 0.0076 & 0.953 & 0.0009 & 0.0082 & 0.0070 & 0.0082 & 0.963 & 0.0016 & 0.0082 & 0.0073 & 0.0084 & 0.952 \\
 & $\sigma_{\epsilon}$ & 0.0094 & 0.0920 & 0.0691 & 0.0924 & 0.946 & 0.0068 & 0.0793 & 0.0642 & 0.0796 & 0.954 & 0.0088 & 0.0924 & 0.0693 & 0.0928 & 0.945 & 0.0139 & 0.1109 & 0.0782 & 0.1118 & 0.940 \\
 & $\alpha$ & 0.0005 & 0.0030 & 0.0027 & 0.0030 & 0.927 & 0.0014 & 0.0025 & 0.0026 & 0.0029 & 0.922 & 0.0007 & 0.0019 & 0.0018 & 0.0020 & 0.935 & 0.0005 & 0.0012 & 0.0012 & 0.0013 & 0.937 \\
 & $\gamma$ & -0.0041 & 0.0938 & 0.0932 & 0.0939 & 0.950 & 0.0275 & 0.0986 & 0.0998 & 0.1024 & 0.947 & 0.0142 & 0.0731 & 0.0716 & 0.0744 & 0.941 & 0.0131 & 0.0523 & 0.0528 & 0.0539 & 0.946 \\
\addlinespace
4 & $\beta_0$ & 0.0276 & 0.4793 & 0.4794 & 0.4799 & 0.948 & 0.0084 & 0.4720 & 0.4804 & 0.4718 & 0.961 & 0.0075 & 0.4773 & 0.4776 & 0.4771 & 0.960 & -0.0345 & 0.4812 & 0.4771 & 0.4822 & 0.944 \\
 & $\beta_1$ & -0.0004 & 0.0121 & 0.0120 & 0.0121 & 0.944 & -0.0001 & 0.0121 & 0.0117 & 0.0121 & 0.947 & 0.0001 & 0.0108 & 0.0112 & 0.0108 & 0.971 & -0.0003 & 0.0107 & 0.0105 & 0.0106 & 0.943 \\
 & $\beta_2$ & -0.0001 & 0.0194 & 0.0193 & 0.0194 & 0.942 & 0.0001 & 0.0196 & 0.0192 & 0.0196 & 0.948 & 0.0002 & 0.0183 & 0.0189 & 0.0183 & 0.963 & -0.0004 & 0.0190 & 0.0186 & 0.0190 & 0.944 \\
 & $\sigma_{b0}$ & 0.0088 & 0.3566 & 0.3638 & 0.3565 & 0.967 & 0.0187 & 0.3637 & 0.3609 & 0.3640 & 0.945 & -0.0016 & 0.3838 & 0.3680 & 0.3836 & 0.948 & -0.0060 & 0.3961 & 0.3654 & 0.3959 & 0.946 \\
 & $\sigma_{b1}$ & 0.0004 & 0.0080 & 0.0080 & 0.0080 & 0.951 & 0.0006 & 0.0077 & 0.0079 & 0.0077 & 0.953 & 0.0003 & 0.0091 & 0.0082 & 0.0091 & 0.959 & 0.0004 & 0.0085 & 0.0080 & 0.0085 & 0.958 \\
 & $\sigma_{\epsilon}$ & 0.0007 & 0.0689 & 0.0676 & 0.0688 & 0.950 & -0.0005 & 0.0648 & 0.0663 & 0.0648 & 0.954 & 0.0017 & 0.0756 & 0.0681 & 0.0756 & 0.944 & 0.0032 & 0.0733 & 0.0658 & 0.0734 & 0.953 \\
 & $\alpha$ & 0.0007 & 0.0027 & 0.0027 & 0.0028 & 0.946 & 0.0008 & 0.0023 & 0.0023 & 0.0024 & 0.933 & 0.0002 & 0.0016 & 0.0016 & 0.0016 & 0.942 & 0.0000 & 0.0009 & 0.0010 & 0.0009 & 0.962 \\
 & $\gamma$ & -0.0044 & 0.0916 & 0.0887 & 0.0916 & 0.947 & -0.0091 & 0.0769 & 0.0783 & 0.0774 & 0.954 & -0.0064 & 0.0547 & 0.0519 & 0.0550 & 0.935 & -0.0029 & 0.0316 & 0.0312 & 0.0317 & 0.949 \\
\addlinespace
5 & $\beta_0$ & 0.0424 & 0.4939 & 0.4795 & 0.4955 & 0.940 & -0.0078 & 0.4733 & 0.4800 & 0.4731 & 0.944 & 0.0042 & 0.4918 & 0.4773 & 0.4915 & 0.943 & -0.0176 & 0.4664 & 0.4767 & 0.4665 & 0.959 \\
 & $\beta_1$ & 0.0002 & 0.0121 & 0.0119 & 0.0121 & 0.951 & 0.0002 & 0.0122 & 0.0117 & 0.0122 & 0.943 & 0.0003 & 0.0107 & 0.0111 & 0.0107 & 0.956 & -0.0003 & 0.0108 & 0.0105 & 0.0108 & 0.949 \\
 & $\beta_2$ & -0.0011 & 0.0197 & 0.0194 & 0.0197 & 0.946 & 0.0003 & 0.0190 & 0.0193 & 0.0190 & 0.951 & -0.0005 & 0.0185 & 0.0191 & 0.0185 & 0.955 & -0.0015 & 0.0192 & 0.0188 & 0.0193 & 0.945 \\
 & $\sigma_{b0}$ & 0.0149 & 0.3521 & 0.3597 & 0.3523 & 0.949 & -0.0164 & 0.3661 & 0.3604 & 0.3663 & 0.941 & -0.0063 & 0.3618 & 0.3565 & 0.3617 & 0.942 & 0.0048 & 0.3705 & 0.3607 & 0.3703 & 0.938 \\
 & $\sigma_{b1}$ & 0.0007 & 0.0080 & 0.0080 & 0.0080 & 0.947 & 0.0003 & 0.0080 & 0.0079 & 0.0080 & 0.942 & 0.0002 & 0.0078 & 0.0078 & 0.0078 & 0.949 & 0.0006 & 0.0091 & 0.0081 & 0.0092 & 0.949 \\
 & $\sigma_{\epsilon}$ & -0.0029 & 0.0665 & 0.0665 & 0.0665 & 0.950 & -0.0027 & 0.0699 & 0.0664 & 0.0699 & 0.933 & 0.0017 & 0.0644 & 0.0648 & 0.0644 & 0.950 & 0.0036 & 0.0721 & 0.0660 & 0.0722 & 0.936 \\
 & $\alpha$ & 0.0006 & 0.0027 & 0.0027 & 0.0028 & 0.943 & 0.0009 & 0.0024 & 0.0023 & 0.0025 & 0.925 & 0.0004 & 0.0016 & 0.0016 & 0.0016 & 0.944 & 0.0000 & 0.0010 & 0.0010 & 0.0010 & 0.959 \\
 & $\gamma$ & -0.0063 & 0.0871 & 0.0886 & 0.0872 & 0.953 & -0.0100 & 0.0820 & 0.0779 & 0.0825 & 0.927 & -0.0072 & 0.0501 & 0.0518 & 0.0505 & 0.962 & -0.0022 & 0.0306 & 0.0310 & 0.0307 & 0.953 \\
\bottomrule
\end{tabular}%
\vspace{0.5ex}
\begin{minipage}{0.95\linewidth}
\footnotesize Notes: $\text{CM}_1$ = administrative censoring. LL1.20 = log-logistic(1.20, 23.0); W0.90, W1.30, and W2.10 = Weibull(shape, 38.0). ESD = empirical SD; PSD = posterior SD; CP = coverage probability. Values summarise the 1000 simulation replicates within each scenario and data-generation distribution.
\end{minipage}
\end{table}

\end{landscape}

\begin{landscape}
\begin{table}[!htbp]
\centering
\tiny
\caption{Detailed sAFT-JM performance by data-generation distribution under controlled 50\% censoring.}
\label{tab:saft-bp-distribution-details-cm2}
\begin{tabular}{llrrrrrrrrrrrrrrrrrrrr}
\toprule
Scen. & Par. & \multicolumn{5}{c}{LL1.20} & \multicolumn{5}{c}{W0.90} & \multicolumn{5}{c}{W1.30} & \multicolumn{5}{c}{W2.10}
\\
\cmidrule(lr){3-7} \cmidrule(lr){8-12} \cmidrule(lr){13-17} \cmidrule(lr){18-22}
 &  & Bias & ESD & PSD & RMSE & CP & Bias & ESD & PSD & RMSE & CP & Bias & ESD & PSD & RMSE & CP & Bias & ESD & PSD & RMSE & CP
\\
\midrule
1 & $\beta_0$ & 0.0067 & 0.4795 & 0.4804 & 0.4793 & 0.949 & 0.0036 & 0.4768 & 0.4817 & 0.4765 & 0.951 & -0.0150 & 0.5036 & 0.4783 & 0.5036 & 0.934 & 0.0024 & 0.4802 & 0.4765 & 0.4800 & 0.947 \\
 & $\beta_1$ & -0.0004 & 0.0146 & 0.0144 & 0.0146 & 0.949 & -0.0005 & 0.0142 & 0.0141 & 0.0142 & 0.944 & -0.0005 & 0.0130 & 0.0131 & 0.0131 & 0.959 & -0.0006 & 0.0122 & 0.0121 & 0.0122 & 0.944 \\
 & $\beta_2$ & 0.0005 & 0.0200 & 0.0200 & 0.0199 & 0.941 & 0.0001 & 0.0193 & 0.0197 & 0.0193 & 0.948 & -0.0002 & 0.0184 & 0.0183 & 0.0184 & 0.952 & 0.0002 & 0.0172 & 0.0169 & 0.0172 & 0.948 \\
 & $\sigma_{b0}$ & -0.0116 & 0.3719 & 0.3620 & 0.3719 & 0.941 & 0.0081 & 0.3513 & 0.3659 & 0.3512 & 0.963 & 0.0026 & 0.3688 & 0.3699 & 0.3687 & 0.952 & 0.0096 & 0.3681 & 0.3658 & 0.3681 & 0.959 \\
 & $\sigma_{b1}$ & -0.0001 & 0.0091 & 0.0090 & 0.0091 & 0.951 & 0.0006 & 0.0091 & 0.0089 & 0.0091 & 0.952 & 0.0007 & 0.0098 & 0.0085 & 0.0098 & 0.946 & 0.0003 & 0.0086 & 0.0079 & 0.0086 & 0.950 \\
 & $\sigma_{\epsilon}$ & 0.0036 & 0.0701 & 0.0714 & 0.0701 & 0.952 & -0.0009 & 0.0743 & 0.0716 & 0.0743 & 0.949 & 0.0016 & 0.0768 & 0.0698 & 0.0767 & 0.953 & 0.0053 & 0.0740 & 0.0667 & 0.0741 & 0.957 \\
 & $\alpha$ & 0.0007 & 0.0033 & 0.0032 & 0.0034 & 0.944 & 0.0018 & 0.0031 & 0.0031 & 0.0036 & 0.920 & 0.0010 & 0.0021 & 0.0021 & 0.0023 & 0.935 & 0.0005 & 0.0014 & 0.0014 & 0.0014 & 0.929 \\
 & $\gamma$ & -0.0035 & 0.1013 & 0.1001 & 0.1013 & 0.946 & -0.0012 & 0.1080 & 0.0983 & 0.1079 & 0.925 & -0.0026 & 0.0733 & 0.0681 & 0.0733 & 0.935 & -0.0013 & 0.0432 & 0.0420 & 0.0432 & 0.935 \\
\addlinespace
2 & $\beta_0$ & 0.0183 & 0.4840 & 0.4755 & 0.4841 & 0.950 & -0.0080 & 0.4802 & 0.4753 & 0.4801 & 0.950 & 0.0206 & 0.4712 & 0.4718 & 0.4714 & 0.940 & 0.0061 & 0.4743 & 0.4700 & 0.4741 & 0.948 \\
 & $\beta_1$ & 0.0001 & 0.0126 & 0.0129 & 0.0126 & 0.959 & -0.0007 & 0.0126 & 0.0126 & 0.0126 & 0.946 & -0.0009 & 0.0114 & 0.0116 & 0.0115 & 0.953 & -0.0010 & 0.0113 & 0.0109 & 0.0113 & 0.942 \\
 & $\beta_2$ & -0.0001 & 0.0168 & 0.0168 & 0.0168 & 0.953 & 0.0007 & 0.0171 & 0.0166 & 0.0171 & 0.945 & 0.0006 & 0.0150 & 0.0153 & 0.0150 & 0.950 & 0.0011 & 0.0148 & 0.0143 & 0.0148 & 0.940 \\
 & $\sigma_{b0}$ & -0.0118 & 0.3964 & 0.3722 & 0.3963 & 0.948 & -0.0203 & 0.3744 & 0.3753 & 0.3748 & 0.953 & -0.0225 & 0.4035 & 0.3765 & 0.4040 & 0.951 & -0.0559 & 0.5200 & 0.4507 & 0.5228 & 0.952 \\
 & $\sigma_{b1}$ & 0.0012 & 0.0082 & 0.0073 & 0.0083 & 0.950 & 0.0009 & 0.0082 & 0.0074 & 0.0082 & 0.953 & 0.0007 & 0.0079 & 0.0069 & 0.0079 & 0.946 & 0.0017 & 0.0095 & 0.0078 & 0.0096 & 0.952 \\
 & $\sigma_{\epsilon}$ & 0.0007 & 0.0726 & 0.0647 & 0.0725 & 0.947 & 0.0061 & 0.0782 & 0.0662 & 0.0784 & 0.947 & 0.0097 & 0.0804 & 0.0634 & 0.0810 & 0.948 & 0.0193 & 0.1228 & 0.0827 & 0.1242 & 0.945 \\
 & $\alpha$ & 0.0006 & 0.0031 & 0.0030 & 0.0032 & 0.936 & 0.0016 & 0.0028 & 0.0029 & 0.0033 & 0.926 & 0.0009 & 0.0020 & 0.0020 & 0.0022 & 0.935 & 0.0005 & 0.0014 & 0.0013 & 0.0015 & 0.924 \\
 & $\gamma$ & -0.0023 & 0.1001 & 0.1001 & 0.1001 & 0.949 & 0.0210 & 0.1073 & 0.1066 & 0.1093 & 0.955 & 0.0093 & 0.0759 & 0.0760 & 0.0765 & 0.947 & 0.0117 & 0.0574 & 0.0558 & 0.0585 & 0.945 \\
\addlinespace
3 & $\beta_0$ & -0.0106 & 0.4812 & 0.4763 & 0.4811 & 0.942 & -0.0299 & 0.4908 & 0.4760 & 0.4915 & 0.949 & -0.0038 & 0.4825 & 0.4733 & 0.4823 & 0.951 & 0.0178 & 0.4571 & 0.4709 & 0.4572 & 0.957 \\
 & $\beta_1$ & -0.0005 & 0.0131 & 0.0130 & 0.0131 & 0.946 & -0.0011 & 0.0131 & 0.0127 & 0.0131 & 0.949 & -0.0007 & 0.0117 & 0.0118 & 0.0117 & 0.948 & -0.0003 & 0.0107 & 0.0108 & 0.0107 & 0.951 \\
 & $\beta_2$ & -0.0000 & 0.0169 & 0.0169 & 0.0168 & 0.951 & 0.0005 & 0.0171 & 0.0167 & 0.0171 & 0.942 & 0.0008 & 0.0153 & 0.0154 & 0.0153 & 0.942 & 0.0002 & 0.0143 & 0.0143 & 0.0143 & 0.954 \\
 & $\sigma_{b0}$ & 0.0138 & 0.3846 & 0.3669 & 0.3847 & 0.938 & 0.0004 & 0.4016 & 0.3753 & 0.4014 & 0.946 & -0.0235 & 0.4515 & 0.3979 & 0.4519 & 0.956 & -0.0334 & 0.4644 & 0.4109 & 0.4654 & 0.953 \\
 & $\sigma_{b1}$ & 0.0004 & 0.0079 & 0.0073 & 0.0079 & 0.946 & 0.0010 & 0.0082 & 0.0074 & 0.0082 & 0.945 & 0.0010 & 0.0084 & 0.0072 & 0.0084 & 0.957 & 0.0010 & 0.0080 & 0.0069 & 0.0080 & 0.956 \\
 & $\sigma_{\epsilon}$ & 0.0029 & 0.0684 & 0.0641 & 0.0685 & 0.959 & 0.0053 & 0.0782 & 0.0662 & 0.0784 & 0.946 & 0.0115 & 0.0870 & 0.0678 & 0.0877 & 0.949 & 0.0105 & 0.0955 & 0.0697 & 0.0961 & 0.951 \\
 & $\alpha$ & 0.0006 & 0.0032 & 0.0030 & 0.0033 & 0.926 & 0.0018 & 0.0028 & 0.0029 & 0.0033 & 0.921 & 0.0008 & 0.0021 & 0.0020 & 0.0022 & 0.931 & 0.0005 & 0.0013 & 0.0013 & 0.0014 & 0.947 \\
 & $\gamma$ & -0.0020 & 0.1003 & 0.1003 & 0.1003 & 0.957 & 0.0265 & 0.1076 & 0.1069 & 0.1108 & 0.942 & 0.0179 & 0.0781 & 0.0765 & 0.0801 & 0.941 & 0.0123 & 0.0559 & 0.0554 & 0.0572 & 0.948 \\
\addlinespace
4 & $\beta_0$ & 0.0241 & 0.4952 & 0.4879 & 0.4955 & 0.942 & -0.0186 & 0.4911 & 0.4881 & 0.4912 & 0.953 & -0.0246 & 0.4751 & 0.4848 & 0.4755 & 0.958 & 0.0064 & 0.4759 & 0.4826 & 0.4757 & 0.950 \\
 & $\beta_1$ & -0.0014 & 0.0164 & 0.0164 & 0.0164 & 0.949 & -0.0011 & 0.0155 & 0.0162 & 0.0155 & 0.950 & 0.0006 & 0.0146 & 0.0149 & 0.0146 & 0.946 & -0.0005 & 0.0137 & 0.0136 & 0.0137 & 0.942 \\
 & $\beta_2$ & -0.0007 & 0.0254 & 0.0260 & 0.0254 & 0.956 & 0.0009 & 0.0260 & 0.0259 & 0.0260 & 0.947 & -0.0014 & 0.0248 & 0.0245 & 0.0249 & 0.947 & 0.0001 & 0.0233 & 0.0232 & 0.0233 & 0.939 \\
 & $\sigma_{b0}$ & 0.0076 & 0.3693 & 0.3718 & 0.3692 & 0.947 & 0.0138 & 0.3705 & 0.3734 & 0.3706 & 0.954 & -0.0044 & 0.3581 & 0.3687 & 0.3579 & 0.963 & -0.0077 & 0.3736 & 0.3686 & 0.3735 & 0.949 \\
 & $\sigma_{b1}$ & 0.0006 & 0.0121 & 0.0122 & 0.0121 & 0.955 & 0.0014 & 0.0116 & 0.0121 & 0.0117 & 0.958 & 0.0007 & 0.0119 & 0.0115 & 0.0120 & 0.941 & 0.0005 & 0.0116 & 0.0110 & 0.0117 & 0.951 \\
 & $\sigma_{\epsilon}$ & -0.0019 & 0.0835 & 0.0852 & 0.0835 & 0.958 & 0.0002 & 0.0845 & 0.0851 & 0.0845 & 0.949 & 0.0062 & 0.0823 & 0.0818 & 0.0825 & 0.959 & -0.0008 & 0.0815 & 0.0787 & 0.0814 & 0.949 \\
 & $\alpha$ & 0.0009 & 0.0035 & 0.0035 & 0.0036 & 0.937 & 0.0012 & 0.0032 & 0.0032 & 0.0034 & 0.935 & 0.0003 & 0.0022 & 0.0022 & 0.0022 & 0.942 & 0.0001 & 0.0014 & 0.0014 & 0.0014 & 0.952 \\
 & $\gamma$ & -0.0080 & 0.1037 & 0.1037 & 0.1040 & 0.941 & -0.0139 & 0.0995 & 0.0996 & 0.1004 & 0.947 & -0.0074 & 0.0710 & 0.0694 & 0.0713 & 0.938 & -0.0049 & 0.0445 & 0.0439 & 0.0447 & 0.951 \\
\addlinespace
5 & $\beta_0$ & 0.0055 & 0.4898 & 0.4866 & 0.4895 & 0.948 & -0.0226 & 0.4755 & 0.4883 & 0.4758 & 0.963 & -0.0128 & 0.4791 & 0.4862 & 0.4790 & 0.952 & 0.0221 & 0.4901 & 0.4837 & 0.4904 & 0.943 \\
 & $\beta_1$ & -0.0021 & 0.0169 & 0.0164 & 0.0170 & 0.933 & 0.0006 & 0.0155 & 0.0163 & 0.0155 & 0.957 & 0.0002 & 0.0144 & 0.0149 & 0.0144 & 0.956 & 0.0002 & 0.0138 & 0.0136 & 0.0138 & 0.947 \\
 & $\beta_2$ & 0.0007 & 0.0269 & 0.0262 & 0.0269 & 0.949 & -0.0007 & 0.0259 & 0.0261 & 0.0259 & 0.952 & -0.0010 & 0.0248 & 0.0247 & 0.0249 & 0.942 & -0.0006 & 0.0231 & 0.0234 & 0.0231 & 0.949 \\
 & $\sigma_{b0}$ & -0.0089 & 0.3798 & 0.3716 & 0.3797 & 0.944 & 0.0099 & 0.3831 & 0.3731 & 0.3831 & 0.935 & 0.0351 & 0.3751 & 0.3685 & 0.3765 & 0.956 & 0.0078 & 0.3701 & 0.3693 & 0.3700 & 0.956 \\
 & $\sigma_{b1}$ & 0.0007 & 0.0120 & 0.0122 & 0.0120 & 0.958 & 0.0015 & 0.0120 & 0.0122 & 0.0120 & 0.947 & 0.0008 & 0.0114 & 0.0115 & 0.0114 & 0.948 & 0.0003 & 0.0117 & 0.0110 & 0.0117 & 0.954 \\
 & $\sigma_{\epsilon}$ & -0.0017 & 0.0827 & 0.0854 & 0.0827 & 0.949 & 0.0045 & 0.0897 & 0.0855 & 0.0897 & 0.935 & 0.0024 & 0.0821 & 0.0810 & 0.0821 & 0.948 & 0.0074 & 0.0833 & 0.0790 & 0.0836 & 0.948 \\
 & $\alpha$ & 0.0009 & 0.0037 & 0.0035 & 0.0038 & 0.935 & 0.0008 & 0.0033 & 0.0032 & 0.0034 & 0.947 & 0.0003 & 0.0023 & 0.0022 & 0.0023 & 0.934 & 0.0001 & 0.0014 & 0.0014 & 0.0014 & 0.959 \\
 & $\gamma$ & -0.0115 & 0.1010 & 0.1036 & 0.1016 & 0.957 & -0.0088 & 0.0985 & 0.0994 & 0.0988 & 0.953 & -0.0068 & 0.0700 & 0.0695 & 0.0703 & 0.945 & -0.0053 & 0.0436 & 0.0438 & 0.0439 & 0.954 \\
\bottomrule
\end{tabular}%
\vspace{0.5ex}
\begin{minipage}{0.95\linewidth}
\footnotesize Notes: $\text{CM}_2$ = controlled 50\% censoring. LL1.20 = log-logistic(1.20, 23.0); W0.90, W1.30, and W2.10 = Weibull(shape, 38.0). ESD = empirical SD; PSD = posterior SD; CP = coverage probability. Values summarise the 1000 simulation replicates within each scenario and data-generation distribution.
\end{minipage}
\end{table}

\end{landscape}

\section{Longitudinal Estimates under Log-Logistic Data Generation}
\label{sec:supp-coxjm-comparison}

This supplementary analysis was restricted to the log-logistic data-generation settings and was used to compare longitudinal-side inference under alternative analysis models. The comparison is not intended as a direct comparison of survival-side treatment effects, because AFT and proportional hazards models parameterise covariate effects on different scales.

\begin{table}[!htbp]
\centering
\normalsize
\caption{Longitudinal estimates under log-logistic data generation.}
\label{tab:longitudinal-ll-model-estimates}
\begin{tabular}{llrrrrrrr}
\toprule
Scen. & Par. & True & \multicolumn{2}{c}{LMM} & \multicolumn{2}{c}{sAFT-JM} & \multicolumn{2}{c}{Cox-JM}
\\
\cmidrule(lr){4-5} \cmidrule(lr){6-7} \cmidrule(lr){8-9}
 &  &  & $\text{CM}_1$ & $\text{CM}_2$ & $\text{CM}_1$ & $\text{CM}_2$ & $\text{CM}_1$ & $\text{CM}_2$
\\
\midrule
1 & $\beta_0$ & 73.000 & 73.000 & 73.011 & 72.998 & 73.007 & 73.000 & 73.009 \\
 & $\beta_1$ & -0.040 & -0.035 & -0.035 & -0.040 & -0.040 & -0.039 & -0.039 \\
 & $\beta_2$ & 0.000 & -0.001 & 0.000 & -0.001 & 0.000 & -0.001 & 0.000 \\
 & $\sigma_{b0}$ & 15.000 & 15.009 & 14.978 & 15.020 & 14.988 & 15.025 & 14.992 \\
 & $\sigma_{b1}$ & 0.200 & 0.200 & 0.199 & 0.200 & 0.200 & 0.200 & 0.199 \\
 & $\sigma_{\epsilon}$ & 12.000 & 11.996 & 12.003 & 11.997 & 12.004 & 11.997 & 12.004 \\
\addlinespace
2 & $\beta_0$ & 73.000 & 72.986 & 73.016 & 72.990 & 73.018 & 72.988 & 73.017 \\
 & $\beta_1$ & -0.040 & -0.035 & -0.034 & -0.040 & -0.040 & -0.038 & -0.038 \\
 & $\beta_2$ & 0.040 & 0.037 & 0.037 & 0.040 & 0.040 & 0.038 & 0.039 \\
 & $\sigma_{b0}$ & 15.000 & 14.992 & 14.988 & 14.975 & 14.988 & 15.006 & 15.003 \\
 & $\sigma_{b1}$ & 0.200 & 0.200 & 0.200 & 0.201 & 0.201 & 0.200 & 0.201 \\
 & $\sigma_{\epsilon}$ & 12.000 & 11.999 & 11.998 & 12.007 & 12.001 & 12.000 & 11.998 \\
\addlinespace
3 & $\beta_0$ & 73.000 & 72.999 & 72.990 & 73.001 & 72.989 & 73.004 & 72.993 \\
 & $\beta_1$ & -0.040 & -0.034 & -0.035 & -0.040 & -0.040 & -0.039 & -0.040 \\
 & $\beta_2$ & -0.040 & -0.043 & -0.043 & -0.040 & -0.040 & -0.040 & -0.040 \\
 & $\sigma_{b0}$ & 15.000 & 15.006 & 15.010 & 14.984 & 15.014 & 15.021 & 15.024 \\
 & $\sigma_{b1}$ & 0.200 & 0.200 & 0.200 & 0.201 & 0.200 & 0.200 & 0.200 \\
 & $\sigma_{\epsilon}$ & 12.000 & 12.001 & 12.001 & 12.009 & 12.003 & 12.001 & 12.001 \\
\addlinespace
4 & $\beta_0$ & 73.000 & 73.044 & 73.039 & 73.028 & 73.024 & 73.031 & 73.025 \\
 & $\beta_1$ & -0.040 & -0.035 & -0.036 & -0.040 & -0.041 & -0.039 & -0.040 \\
 & $\beta_2$ & 0.040 & 0.044 & 0.044 & 0.040 & 0.039 & 0.041 & 0.040 \\
 & $\sigma_{b0}$ & 15.000 & 14.998 & 15.002 & 15.009 & 15.008 & 15.016 & 15.012 \\
 & $\sigma_{b1}$ & 0.200 & 0.200 & 0.200 & 0.200 & 0.201 & 0.200 & 0.200 \\
 & $\sigma_{\epsilon}$ & 12.000 & 12.000 & 11.997 & 12.001 & 11.998 & 12.000 & 11.999 \\
\addlinespace
5 & $\beta_0$ & 73.000 & 73.060 & 73.022 & 73.042 & 73.005 & 73.049 & 73.010 \\
 & $\beta_1$ & -0.040 & -0.034 & -0.036 & -0.040 & -0.042 & -0.038 & -0.041 \\
 & $\beta_2$ & -0.040 & -0.037 & -0.034 & -0.041 & -0.039 & -0.040 & -0.038 \\
 & $\sigma_{b0}$ & 15.000 & 15.001 & 14.986 & 15.015 & 14.991 & 15.020 & 14.995 \\
 & $\sigma_{b1}$ & 0.200 & 0.200 & 0.200 & 0.201 & 0.201 & 0.200 & 0.200 \\
 & $\sigma_{\epsilon}$ & 12.000 & 11.997 & 11.997 & 11.997 & 11.998 & 11.997 & 11.999 \\
\bottomrule
\end{tabular}
\vspace{0.5ex}
\begin{minipage}{0.95\linewidth}
\footnotesize Notes: $\text{CM}_1$ = administrative censoring; $\text{CM}_2$ = controlled 50\% censoring. Only log-logistic data-generation configurations are included. Entries are means of point estimates across simulations.
\end{minipage}
\end{table}

\begin{landscape}
\begin{figure}[p]
\centering
\includegraphics[width=\linewidth, height=\textheight, keepaspectratio]{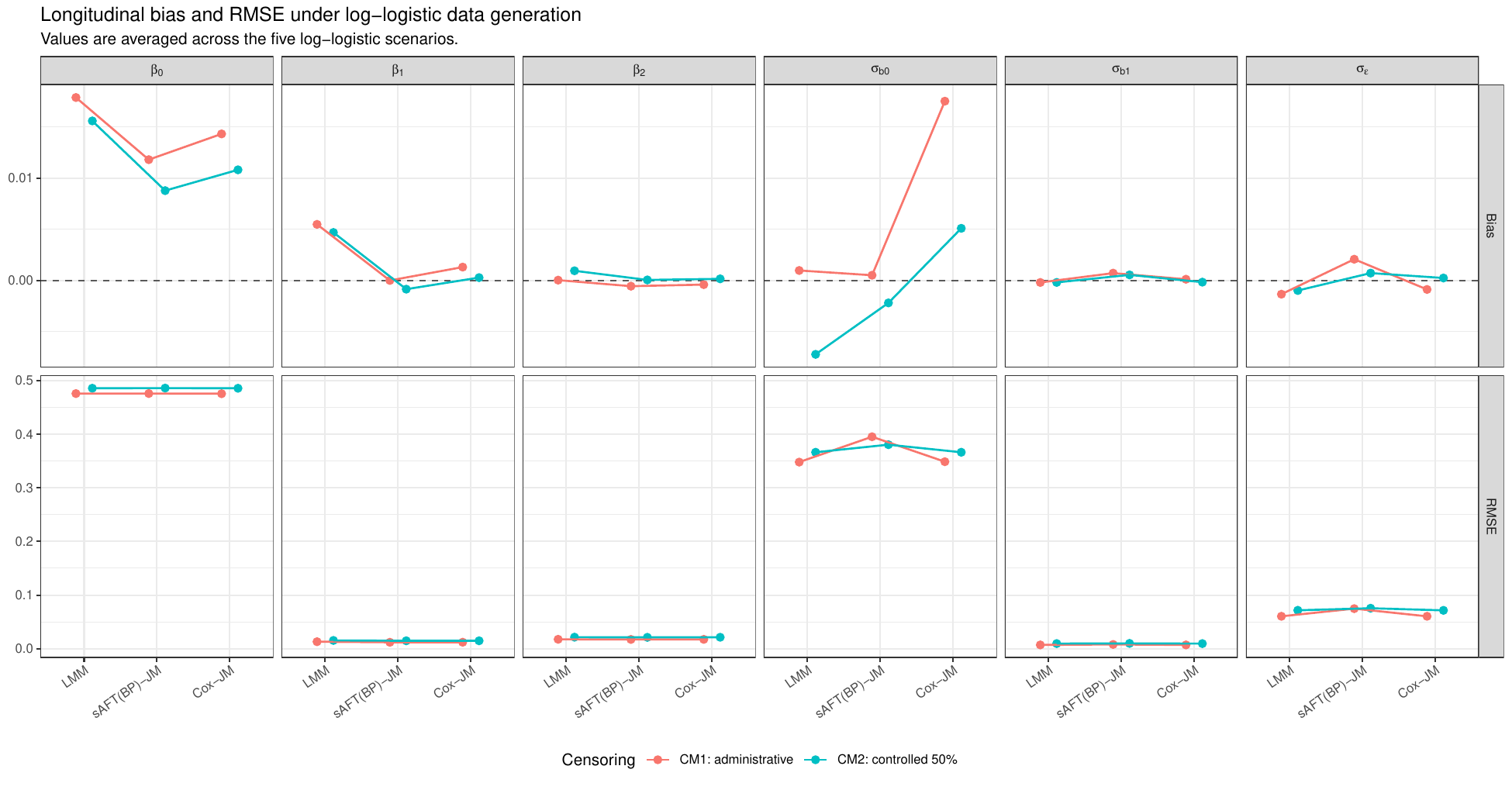}
\caption{Bias and RMSE of LMM, sAFT-JM and Cox-JM's longitudinal coefficient estimates under log-logistic data generation. Values are averaged across the five scenarios of Table 2 in the main article, each dot therefore summarises 5000 simulation replicates.}
\label{fig:longitudinal_LL_bias_rmse}
\end{figure}
\end{landscape}

\begin{landscape}
\begin{table}[!htbp]
\centering
\small
\caption{Detailed longitudinal performance under log-logistic data generation and administrative censoring.}
\label{tab:longitudinal-ll-model-details-cm1}
\begin{tabular}{llrrrrrrrrrrrrrrr}
\toprule
Scen. & Par. & \multicolumn{5}{c}{LMM} & \multicolumn{5}{c}{sAFT-JM} & \multicolumn{5}{c}{Cox-JM}
\\
\cmidrule(lr){3-7} \cmidrule(lr){8-12} \cmidrule(lr){13-17}
 &  & Bias & ESD & PSD & RMSE & CP & Bias & ESD & PSD & RMSE & CP & Bias & ESD & PSD & RMSE & CP
\\
\midrule
1 & $\beta_0$ & -0.0003 & 0.4574 & 0.4763 & 0.4572 & 0.969 & -0.0022 & 0.4573 & 0.4759 & 0.4571 & 0.969 & 0.0002 & 0.4579 & 0.4769 & 0.4576 & 0.968 \\
 & $\beta_1$ & 0.0054 & 0.0113 & 0.0118 & 0.0126 & 0.939 & 0.0000 & 0.0114 & 0.0120 & 0.0114 & 0.959 & 0.0014 & 0.0115 & 0.0118 & 0.0115 & 0.953 \\
 & $\beta_2$ & -0.0008 & 0.0165 & 0.0166 & 0.0165 & 0.948 & -0.0008 & 0.0165 & 0.0166 & 0.0165 & 0.947 & -0.0010 & 0.0166 & 0.0163 & 0.0166 & 0.941 \\
 & $\sigma_{b0}$ & 0.0086 & 0.3458 & -- & 0.3458 & -- & 0.0199 & 0.3471 & 0.3580 & 0.3475 & 0.954 & 0.0247 & 0.3461 & 0.3785 & 0.3468 & 0.971 \\
 & $\sigma_{b1}$ & -0.0002 & 0.0070 & -- & 0.0070 & -- & 0.0005 & 0.0071 & 0.0070 & 0.0071 & 0.947 & 0.0001 & 0.0070 & 0.0068 & 0.0070 & 0.940 \\
 & $\sigma_{\epsilon}$ & -0.0038 & 0.0595 & -- & 0.0596 & -- & -0.0034 & 0.0598 & 0.0608 & 0.0598 & 0.950 & -0.0033 & 0.0595 & 0.0603 & 0.0595 & 0.949 \\
\addlinespace
2 & $\beta_0$ & -0.0141 & 0.4764 & 0.4735 & 0.4764 & 0.948 & -0.0095 & 0.4783 & 0.4726 & 0.4781 & 0.945 & -0.0124 & 0.4763 & 0.4739 & 0.4762 & 0.946 \\
 & $\beta_1$ & 0.0055 & 0.0120 & 0.0118 & 0.0132 & 0.925 & -0.0000 & 0.0121 & 0.0120 & 0.0121 & 0.952 & 0.0016 & 0.0120 & 0.0118 & 0.0121 & 0.945 \\
 & $\beta_2$ & -0.0033 & 0.0152 & 0.0155 & 0.0156 & 0.943 & -0.0003 & 0.0153 & 0.0156 & 0.0153 & 0.954 & -0.0016 & 0.0152 & 0.0153 & 0.0153 & 0.949 \\
 & $\sigma_{b0}$ & -0.0083 & 0.3479 & -- & 0.3478 & -- & -0.0247 & 0.4512 & 0.3959 & 0.4517 & 0.950 & 0.0059 & 0.3481 & 0.3735 & 0.3480 & 0.965 \\
 & $\sigma_{b1}$ & 0.0000 & 0.0063 & -- & 0.0063 & -- & 0.0011 & 0.0082 & 0.0071 & 0.0083 & 0.952 & 0.0003 & 0.0063 & 0.0061 & 0.0063 & 0.954 \\
 & $\sigma_{\epsilon}$ & -0.0006 & 0.0539 & -- & 0.0538 & -- & 0.0067 & 0.0857 & 0.0674 & 0.0859 & 0.961 & -0.0001 & 0.0540 & 0.0559 & 0.0539 & 0.959 \\
\addlinespace
3 & $\beta_0$ & -0.0010 & 0.4671 & 0.4739 & 0.4669 & 0.955 & 0.0007 & 0.4684 & 0.4735 & 0.4681 & 0.950 & 0.0040 & 0.4669 & 0.4743 & 0.4667 & 0.956 \\
 & $\beta_1$ & 0.0057 & 0.0119 & 0.0118 & 0.0132 & 0.926 & 0.0002 & 0.0120 & 0.0120 & 0.0120 & 0.947 & 0.0006 & 0.0120 & 0.0118 & 0.0120 & 0.945 \\
 & $\beta_2$ & -0.0033 & 0.0154 & 0.0155 & 0.0158 & 0.938 & -0.0005 & 0.0155 & 0.0156 & 0.0155 & 0.952 & 0.0000 & 0.0156 & 0.0152 & 0.0156 & 0.933 \\
 & $\sigma_{b0}$ & 0.0059 & 0.3486 & -- & 0.3485 & -- & -0.0162 & 0.4678 & 0.4044 & 0.4678 & 0.943 & 0.0207 & 0.3493 & 0.3736 & 0.3498 & 0.954 \\
 & $\sigma_{b1}$ & -0.0002 & 0.0063 & -- & 0.0063 & -- & 0.0010 & 0.0083 & 0.0072 & 0.0083 & 0.945 & 0.0001 & 0.0063 & 0.0061 & 0.0063 & 0.940 \\
 & $\sigma_{\epsilon}$ & 0.0011 & 0.0571 & -- & 0.0571 & -- & 0.0094 & 0.0920 & 0.0691 & 0.0924 & 0.946 & 0.0015 & 0.0572 & 0.0560 & 0.0572 & 0.945 \\
\addlinespace
4 & $\beta_0$ & 0.0443 & 0.4793 & 0.4798 & 0.4811 & 0.949 & 0.0276 & 0.4793 & 0.4794 & 0.4799 & 0.948 & 0.0310 & 0.4793 & 0.4803 & 0.4800 & 0.948 \\
 & $\beta_1$ & 0.0051 & 0.0120 & 0.0118 & 0.0130 & 0.918 & -0.0004 & 0.0121 & 0.0120 & 0.0121 & 0.944 & 0.0012 & 0.0121 & 0.0118 & 0.0122 & 0.939 \\
 & $\beta_2$ & 0.0042 & 0.0193 & 0.0192 & 0.0198 & 0.937 & -0.0001 & 0.0194 & 0.0193 & 0.0194 & 0.942 & 0.0005 & 0.0195 & 0.0190 & 0.0195 & 0.937 \\
 & $\sigma_{b0}$ & -0.0023 & 0.3468 & -- & 0.3466 & -- & 0.0088 & 0.3566 & 0.3638 & 0.3565 & 0.967 & 0.0161 & 0.3471 & 0.3841 & 0.3473 & 0.979 \\
 & $\sigma_{b1}$ & -0.0004 & 0.0077 & -- & 0.0077 & -- & 0.0004 & 0.0080 & 0.0080 & 0.0080 & 0.951 & -0.0001 & 0.0077 & 0.0078 & 0.0077 & 0.954 \\
 & $\sigma_{\epsilon}$ & -0.0003 & 0.0651 & -- & 0.0650 & -- & 0.0007 & 0.0689 & 0.0676 & 0.0688 & 0.950 & 0.0002 & 0.0650 & 0.0664 & 0.0650 & 0.953 \\
\addlinespace
5 & $\beta_0$ & 0.0604 & 0.4934 & 0.4800 & 0.4968 & 0.937 & 0.0424 & 0.4939 & 0.4795 & 0.4955 & 0.940 & 0.0487 & 0.4944 & 0.4805 & 0.4965 & 0.936 \\
 & $\beta_1$ & 0.0056 & 0.0120 & 0.0118 & 0.0132 & 0.917 & 0.0002 & 0.0121 & 0.0119 & 0.0121 & 0.951 & 0.0018 & 0.0122 & 0.0118 & 0.0123 & 0.940 \\
 & $\beta_2$ & 0.0033 & 0.0197 & 0.0193 & 0.0200 & 0.949 & -0.0011 & 0.0197 & 0.0194 & 0.0197 & 0.946 & 0.0001 & 0.0199 & 0.0190 & 0.0199 & 0.941 \\
 & $\sigma_{b0}$ & 0.0010 & 0.3511 & -- & 0.3509 & -- & 0.0149 & 0.3521 & 0.3597 & 0.3523 & 0.949 & 0.0201 & 0.3510 & 0.3846 & 0.3514 & 0.960 \\
 & $\sigma_{b1}$ & -0.0001 & 0.0080 & -- & 0.0079 & -- & 0.0007 & 0.0080 & 0.0080 & 0.0080 & 0.947 & 0.0002 & 0.0080 & 0.0078 & 0.0080 & 0.946 \\
 & $\sigma_{\epsilon}$ & -0.0032 & 0.0664 & -- & 0.0664 & -- & -0.0029 & 0.0665 & 0.0665 & 0.0665 & 0.950 & -0.0026 & 0.0663 & 0.0666 & 0.0664 & 0.955 \\
\bottomrule
\end{tabular}%
\vspace{0.5ex}
\begin{minipage}{0.95\linewidth}
\footnotesize Notes: $\text{CM}_1$ = administrative censoring. Only log-logistic data-generation configurations are included. ESD = empirical SD; PSD = posterior SD; CP = coverage probability. Values summarise the 1000 simulation replicates within each scenario and fitted model. For LMM variance parameters, unavailable interval-based Bayesian summaries remain --.
\end{minipage}
\end{table}

\end{landscape}

\begin{landscape}
\begin{table}[!htbp]
\centering
\small
\caption{Detailed longitudinal performance under log-logistic data generation and controlled 50\% censoring.}
\label{tab:longitudinal-ll-model-details-cm2}
\begin{tabular}{llrrrrrrrrrrrrrrr}
\toprule
Scen. & Par. & \multicolumn{5}{c}{LMM} & \multicolumn{5}{c}{sAFT-JM} & \multicolumn{5}{c}{Cox-JM}
\\
\cmidrule(lr){3-7} \cmidrule(lr){8-12} \cmidrule(lr){13-17}
 &  & Bias & ESD & PSD & RMSE & CP & Bias & ESD & PSD & RMSE & CP & Bias & ESD & PSD & RMSE & CP
\\
\midrule
1 & $\beta_0$ & 0.0111 & 0.4780 & 0.4809 & 0.4779 & 0.948 & 0.0067 & 0.4795 & 0.4804 & 0.4793 & 0.949 & 0.0089 & 0.4782 & 0.4815 & 0.4780 & 0.947 \\
 & $\beta_1$ & 0.0050 & 0.0146 & 0.0143 & 0.0155 & 0.934 & -0.0004 & 0.0146 & 0.0144 & 0.0146 & 0.949 & 0.0008 & 0.0147 & 0.0142 & 0.0147 & 0.939 \\
 & $\beta_2$ & 0.0005 & 0.0199 & 0.0200 & 0.0199 & 0.937 & 0.0005 & 0.0200 & 0.0200 & 0.0199 & 0.941 & 0.0004 & 0.0203 & 0.0197 & 0.0203 & 0.933 \\
 & $\sigma_{b0}$ & -0.0223 & 0.3713 & -- & 0.3718 & -- & -0.0116 & 0.3719 & 0.3620 & 0.3719 & 0.941 & -0.0080 & 0.3715 & 0.3872 & 0.3714 & 0.961 \\
 & $\sigma_{b1}$ & -0.0009 & 0.0090 & -- & 0.0091 & -- & -0.0001 & 0.0091 & 0.0090 & 0.0091 & 0.951 & -0.0006 & 0.0091 & 0.0088 & 0.0091 & 0.942 \\
 & $\sigma_{\epsilon}$ & 0.0031 & 0.0701 & -- & 0.0701 & -- & 0.0036 & 0.0701 & 0.0714 & 0.0701 & 0.952 & 0.0040 & 0.0701 & 0.0713 & 0.0701 & 0.955 \\
\addlinespace
2 & $\beta_0$ & 0.0164 & 0.4840 & 0.4758 & 0.4840 & 0.951 & 0.0183 & 0.4840 & 0.4755 & 0.4841 & 0.950 & 0.0168 & 0.4839 & 0.4764 & 0.4840 & 0.947 \\
 & $\beta_1$ & 0.0056 & 0.0126 & 0.0128 & 0.0138 & 0.929 & 0.0001 & 0.0126 & 0.0129 & 0.0126 & 0.959 & 0.0016 & 0.0126 & 0.0127 & 0.0127 & 0.955 \\
 & $\beta_2$ & -0.0030 & 0.0169 & 0.0167 & 0.0171 & 0.947 & -0.0001 & 0.0168 & 0.0168 & 0.0168 & 0.953 & -0.0012 & 0.0168 & 0.0165 & 0.0169 & 0.936 \\
 & $\sigma_{b0}$ & -0.0116 & 0.3500 & -- & 0.3501 & -- & -0.0118 & 0.3964 & 0.3722 & 0.3963 & 0.948 & 0.0029 & 0.3501 & 0.3781 & 0.3499 & 0.965 \\
 & $\sigma_{b1}$ & 0.0003 & 0.0069 & -- & 0.0069 & -- & 0.0012 & 0.0082 & 0.0073 & 0.0083 & 0.950 & 0.0006 & 0.0070 & 0.0068 & 0.0070 & 0.951 \\
 & $\sigma_{\epsilon}$ & -0.0023 & 0.0616 & -- & 0.0616 & -- & 0.0007 & 0.0726 & 0.0647 & 0.0725 & 0.947 & -0.0018 & 0.0616 & 0.0602 & 0.0616 & 0.945 \\
\addlinespace
3 & $\beta_0$ & -0.0103 & 0.4811 & 0.4767 & 0.4810 & 0.947 & -0.0106 & 0.4812 & 0.4763 & 0.4811 & 0.942 & -0.0067 & 0.4815 & 0.4771 & 0.4813 & 0.946 \\
 & $\beta_1$ & 0.0050 & 0.0130 & 0.0129 & 0.0139 & 0.943 & -0.0005 & 0.0131 & 0.0130 & 0.0131 & 0.946 & 0.0001 & 0.0130 & 0.0128 & 0.0130 & 0.943 \\
 & $\beta_2$ & -0.0029 & 0.0168 & 0.0168 & 0.0171 & 0.945 & -0.0000 & 0.0169 & 0.0169 & 0.0168 & 0.951 & 0.0001 & 0.0169 & 0.0166 & 0.0169 & 0.943 \\
 & $\sigma_{b0}$ & 0.0100 & 0.3615 & -- & 0.3614 & -- & 0.0138 & 0.3846 & 0.3669 & 0.3847 & 0.938 & 0.0239 & 0.3621 & 0.3790 & 0.3627 & 0.960 \\
 & $\sigma_{b1}$ & -0.0004 & 0.0071 & -- & 0.0072 & -- & 0.0004 & 0.0079 & 0.0073 & 0.0079 & 0.946 & -0.0001 & 0.0071 & 0.0069 & 0.0071 & 0.944 \\
 & $\sigma_{\epsilon}$ & 0.0006 & 0.0597 & -- & 0.0597 & -- & 0.0029 & 0.0684 & 0.0641 & 0.0685 & 0.959 & 0.0011 & 0.0597 & 0.0607 & 0.0597 & 0.958 \\
\addlinespace
4 & $\beta_0$ & 0.0387 & 0.4949 & 0.4881 & 0.4961 & 0.948 & 0.0241 & 0.4952 & 0.4879 & 0.4955 & 0.942 & 0.0255 & 0.4952 & 0.4883 & 0.4956 & 0.945 \\
 & $\beta_1$ & 0.0043 & 0.0163 & 0.0162 & 0.0169 & 0.941 & -0.0014 & 0.0164 & 0.0164 & 0.0164 & 0.949 & -0.0002 & 0.0164 & 0.0162 & 0.0164 & 0.945 \\
 & $\beta_2$ & 0.0043 & 0.0254 & 0.0259 & 0.0258 & 0.947 & -0.0007 & 0.0254 & 0.0260 & 0.0254 & 0.956 & -0.0000 & 0.0255 & 0.0257 & 0.0255 & 0.952 \\
 & $\sigma_{b0}$ & 0.0022 & 0.3690 & -- & 0.3689 & -- & 0.0076 & 0.3693 & 0.3718 & 0.3692 & 0.947 & 0.0121 & 0.3691 & 0.3978 & 0.3691 & 0.960 \\
 & $\sigma_{b1}$ & 0.0000 & 0.0124 & -- & 0.0124 & -- & 0.0006 & 0.0121 & 0.0122 & 0.0121 & 0.955 & -0.0003 & 0.0122 & 0.0119 & 0.0122 & 0.945 \\
 & $\sigma_{\epsilon}$ & -0.0032 & 0.0836 & -- & 0.0836 & -- & -0.0019 & 0.0835 & 0.0852 & 0.0835 & 0.958 & -0.0012 & 0.0835 & 0.0852 & 0.0834 & 0.956 \\
\addlinespace
5 & $\beta_0$ & 0.0220 & 0.4888 & 0.4877 & 0.4891 & 0.947 & 0.0055 & 0.4898 & 0.4866 & 0.4895 & 0.948 & 0.0095 & 0.4890 & 0.4880 & 0.4889 & 0.948 \\
 & $\beta_1$ & 0.0035 & 0.0168 & 0.0163 & 0.0171 & 0.934 & -0.0021 & 0.0169 & 0.0164 & 0.0170 & 0.933 & -0.0008 & 0.0170 & 0.0162 & 0.0170 & 0.936 \\
 & $\beta_2$ & 0.0059 & 0.0269 & 0.0260 & 0.0275 & 0.943 & 0.0007 & 0.0269 & 0.0262 & 0.0269 & 0.949 & 0.0016 & 0.0270 & 0.0259 & 0.0270 & 0.937 \\
 & $\sigma_{b0}$ & -0.0142 & 0.3791 & -- & 0.3792 & -- & -0.0089 & 0.3798 & 0.3716 & 0.3797 & 0.944 & -0.0054 & 0.3783 & 0.3977 & 0.3782 & 0.958 \\
 & $\sigma_{b1}$ & 0.0001 & 0.0122 & -- & 0.0122 & -- & 0.0007 & 0.0120 & 0.0122 & 0.0120 & 0.958 & -0.0003 & 0.0121 & 0.0119 & 0.0121 & 0.947 \\
 & $\sigma_{\epsilon}$ & -0.0031 & 0.0829 & -- & 0.0829 & -- & -0.0017 & 0.0827 & 0.0854 & 0.0827 & 0.949 & -0.0010 & 0.0827 & 0.0854 & 0.0827 & 0.951 \\
\bottomrule
\end{tabular}%
\vspace{0.5ex}
\begin{minipage}{0.95\linewidth}
\footnotesize Notes: $\text{CM}_2$ = controlled 50\% censoring. Only log-logistic data-generation configurations are included. ESD = empirical SD; PSD = posterior SD; CP = coverage probability. Values summarise the 1000 simulation replicates within each scenario and fitted model. For LMM variance parameters, unavailable interval-based Bayesian summaries remain --.
\end{minipage}
\end{table}

\end{landscape}

\section{Sensitivity to the Prior Scale for \texorpdfstring{\(\alpha\)}{alpha}}
\label{sec:supp-alpha-sensitivity}

The prior scale for \(\alpha\) was varied to assess whether the main findings were sensitive to the weakly informative prior used in the primary simulation. Across the log-logistic data-generation settings, the estimated longitudinal, survival, and association parameters changed very little across the prior scales, suggesting that the main conclusions were not driven by this prior choice.

\begin{table}[!htbp]
\centering
\footnotesize
\caption{$\alpha$-prior sensitivity estimates in the log-logistic data-generation settings.}
\label{tab:alpha-prior-sensitivity-estimates}
\begin{tabular}{llrrrrrrrrr}
\toprule
Scen. & Par. & True & \multicolumn{2}{c}{$\sigma_{\alpha}=0.114$} & \multicolumn{2}{c}{$\sigma_{\alpha}=0.207$} & \multicolumn{2}{c}{$\sigma_{\alpha}=0.354$} & \multicolumn{2}{c}{$\sigma_{\alpha}=0.561$}
\\
\cmidrule(lr){4-5} \cmidrule(lr){6-7} \cmidrule(lr){8-9} \cmidrule(lr){10-11}
 &  &  & $\text{CM}_1$ & $\text{CM}_2$ & $\text{CM}_1$ & $\text{CM}_2$ & $\text{CM}_1$ & $\text{CM}_2$ & $\text{CM}_1$ & $\text{CM}_2$
\\
\midrule
1 & $\beta_0$ & 73.000 & 72.998 & 73.007 & 72.998 & 73.008 & 72.998 & 73.007 & 72.997 & 73.008 \\
 & $\beta_1$ & -0.040 & -0.040 & -0.040 & -0.040 & -0.040 & -0.040 & -0.040 & -0.040 & -0.040 \\
 & $\beta_2$ & 0.000 & -0.001 & 0.000 & -0.001 & 0.001 & -0.001 & 0.000 & -0.001 & 0.000 \\
 & $\sigma_{b0}$ & 15.000 & 15.011 & 14.988 & 15.016 & 14.985 & 15.020 & 14.988 & 15.016 & 14.988 \\
 & $\sigma_{b1}$ & 0.200 & 0.201 & 0.200 & 0.201 & 0.200 & 0.200 & 0.200 & 0.201 & 0.200 \\
 & $\sigma_{\epsilon}$ & 12.000 & 11.999 & 12.004 & 11.998 & 12.005 & 11.997 & 12.004 & 11.998 & 12.004 \\
 & $\alpha$ & 0.012 & 0.012 & 0.013 & 0.012 & 0.013 & 0.012 & 0.013 & 0.012 & 0.013 \\
 & $\gamma$ & 0.000 & -0.004 & -0.004 & -0.004 & -0.004 & -0.004 & -0.003 & -0.004 & -0.004 \\
\addlinespace
2 & $\beta_0$ & 73.000 & 72.993 & 73.019 & 72.991 & 73.019 & 72.990 & 73.018 & 72.990 & 73.019 \\
 & $\beta_1$ & -0.040 & -0.040 & -0.040 & -0.040 & -0.040 & -0.040 & -0.040 & -0.040 & -0.040 \\
 & $\beta_2$ & 0.040 & 0.040 & 0.040 & 0.040 & 0.040 & 0.040 & 0.040 & 0.040 & 0.040 \\
 & $\sigma_{b0}$ & 15.000 & 14.973 & 14.989 & 14.985 & 14.992 & 14.975 & 14.988 & 14.980 & 14.996 \\
 & $\sigma_{b1}$ & 0.200 & 0.201 & 0.201 & 0.201 & 0.201 & 0.201 & 0.201 & 0.201 & 0.201 \\
 & $\sigma_{\epsilon}$ & 12.000 & 12.007 & 12.001 & 12.004 & 12.000 & 12.007 & 12.001 & 12.006 & 11.999 \\
 & $\alpha$ & 0.012 & 0.012 & 0.013 & 0.012 & 0.013 & 0.012 & 0.013 & 0.012 & 0.013 \\
 & $\gamma$ & 0.900 & 0.892 & 0.898 & 0.892 & 0.898 & 0.892 & 0.898 & 0.892 & 0.898 \\
\addlinespace
3 & $\beta_0$ & 73.000 & 73.000 & 72.991 & 73.000 & 72.989 & 73.001 & 72.989 & 72.999 & 72.989 \\
 & $\beta_1$ & -0.040 & -0.040 & -0.041 & -0.040 & -0.041 & -0.040 & -0.040 & -0.040 & -0.041 \\
 & $\beta_2$ & -0.040 & -0.040 & -0.040 & -0.040 & -0.040 & -0.040 & -0.040 & -0.040 & -0.040 \\
 & $\sigma_{b0}$ & 15.000 & 15.008 & 15.005 & 15.008 & 15.002 & 14.984 & 15.014 & 15.004 & 15.001 \\
 & $\sigma_{b1}$ & 0.200 & 0.201 & 0.201 & 0.201 & 0.201 & 0.201 & 0.200 & 0.201 & 0.201 \\
 & $\sigma_{\epsilon}$ & 12.000 & 12.004 & 12.005 & 12.003 & 12.006 & 12.009 & 12.003 & 12.004 & 12.006 \\
 & $\alpha$ & 0.012 & 0.013 & 0.013 & 0.013 & 0.013 & 0.013 & 0.013 & 0.013 & 0.013 \\
 & $\gamma$ & 0.900 & 0.896 & 0.898 & 0.896 & 0.898 & 0.896 & 0.898 & 0.896 & 0.898 \\
\addlinespace
4 & $\beta_0$ & 73.000 & 73.028 & 73.024 & 73.028 & 73.023 & 73.028 & 73.024 & 73.028 & 73.024 \\
 & $\beta_1$ & -0.040 & -0.040 & -0.041 & -0.040 & -0.041 & -0.040 & -0.041 & -0.040 & -0.041 \\
 & $\beta_2$ & 0.040 & 0.040 & 0.039 & 0.040 & 0.039 & 0.040 & 0.039 & 0.040 & 0.039 \\
 & $\sigma_{b0}$ & 15.000 & 15.005 & 15.008 & 15.012 & 15.008 & 15.009 & 15.008 & 15.012 & 15.008 \\
 & $\sigma_{b1}$ & 0.200 & 0.200 & 0.201 & 0.200 & 0.201 & 0.200 & 0.201 & 0.200 & 0.201 \\
 & $\sigma_{\epsilon}$ & 12.000 & 12.002 & 11.998 & 12.000 & 11.998 & 12.001 & 11.998 & 12.000 & 11.998 \\
 & $\alpha$ & 0.012 & 0.013 & 0.013 & 0.013 & 0.013 & 0.013 & 0.013 & 0.013 & 0.013 \\
 & $\gamma$ & -0.900 & -0.904 & -0.908 & -0.904 & -0.908 & -0.904 & -0.908 & -0.904 & -0.908 \\
\addlinespace
5 & $\beta_0$ & 73.000 & 73.042 & 73.006 & 73.042 & 73.006 & 73.042 & 73.005 & 73.041 & 73.004 \\
 & $\beta_1$ & -0.040 & -0.040 & -0.042 & -0.040 & -0.042 & -0.040 & -0.042 & -0.040 & -0.042 \\
 & $\beta_2$ & -0.040 & -0.041 & -0.039 & -0.041 & -0.039 & -0.041 & -0.039 & -0.041 & -0.039 \\
 & $\sigma_{b0}$ & 15.000 & 15.015 & 14.990 & 15.012 & 14.991 & 15.015 & 14.991 & 15.015 & 14.990 \\
 & $\sigma_{b1}$ & 0.200 & 0.201 & 0.201 & 0.201 & 0.201 & 0.201 & 0.201 & 0.201 & 0.201 \\
 & $\sigma_{\epsilon}$ & 12.000 & 11.997 & 11.998 & 11.998 & 11.998 & 11.997 & 11.998 & 11.997 & 11.998 \\
 & $\alpha$ & 0.012 & 0.013 & 0.013 & 0.013 & 0.013 & 0.013 & 0.013 & 0.013 & 0.013 \\
 & $\gamma$ & -0.900 & -0.906 & -0.911 & -0.906 & -0.911 & -0.906 & -0.911 & -0.906 & -0.911 \\
\bottomrule
\end{tabular}
\vspace{0.5ex}
\begin{minipage}{0.95\linewidth}
\footnotesize Notes: $\sigma_{\alpha}$ denotes the prior SD for $\alpha$, with $\sigma_{\alpha}=\log(F_{\max})/1.96$ for $F_{\max}=1.25,1.5,2,3$. $\text{CM}_1$ = administrative censoring; $\text{CM}_2$ = controlled 50\% censoring. Entries are means of posterior means across simulations.
\end{minipage}
\end{table}

\begin{landscape}
\begin{figure}[p]
\centering
\includegraphics[width=\linewidth, height=\textheight, keepaspectratio]{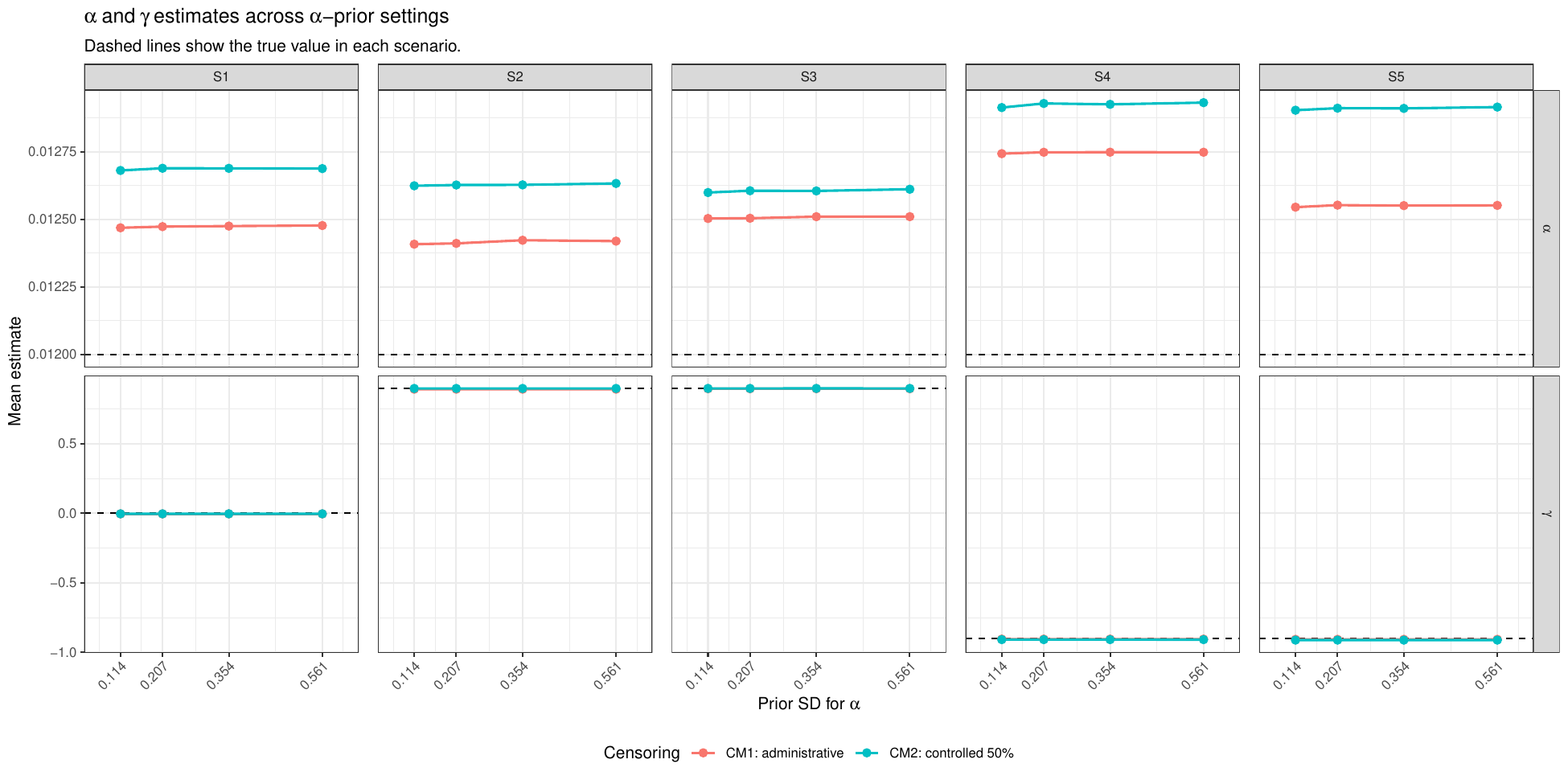}
\caption{$\alpha$-prior sensitivity analysis. $\gamma$ and $\alpha$ estimates over four $\alpha$ prior SD values: 0.114, 0.207, 0.354, and 0.561. Close to horizontal line indicates $\gamma$ and $\alpha$ estimates not sensitive to $\alpha$ prior SD values.}
\label{fig:alpha_prior_gamma_alpha_estimates}
\end{figure}
\end{landscape}

\begin{landscape}
\begin{table}[!htbp]
\centering
\tiny
\caption{Detailed $\alpha$-prior sensitivity statistics for log-logistic generated data under administrative censoring.}
\label{tab:alpha-prior-sensitivity-details-cm1}
\begin{tabular}{llrrrrrrrrrrrrrrrrrrrr}
\toprule
Scen. & Par. & \multicolumn{5}{c}{$\sigma_{\alpha}=0.114$} & \multicolumn{5}{c}{$\sigma_{\alpha}=0.207$} & \multicolumn{5}{c}{$\sigma_{\alpha}=0.354$} & \multicolumn{5}{c}{$\sigma_{\alpha}=0.561$}
\\
\cmidrule(lr){3-7} \cmidrule(lr){8-12} \cmidrule(lr){13-17} \cmidrule(lr){18-22}
 &  & Bias & ESD & PSD & RMSE & CP & Bias & ESD & PSD & RMSE & CP & Bias & ESD & PSD & RMSE & CP & Bias & ESD & PSD & RMSE & CP
\\
\midrule
1 & $\beta_0$ & -0.0019 & 0.4580 & 0.4761 & 0.4578 & 0.966 & -0.0021 & 0.4577 & 0.4757 & 0.4575 & 0.968 & -0.0022 & 0.4573 & 0.4759 & 0.4571 & 0.969 & -0.0026 & 0.4575 & 0.4759 & 0.4572 & 0.967 \\
 & $\beta_1$ & -0.0000 & 0.0114 & 0.0120 & 0.0114 & 0.960 & -0.0000 & 0.0114 & 0.0120 & 0.0114 & 0.958 & 0.0000 & 0.0114 & 0.0120 & 0.0114 & 0.959 & 0.0000 & 0.0114 & 0.0120 & 0.0114 & 0.959 \\
 & $\beta_2$ & -0.0008 & 0.0165 & 0.0167 & 0.0165 & 0.949 & -0.0008 & 0.0165 & 0.0166 & 0.0165 & 0.948 & -0.0008 & 0.0165 & 0.0166 & 0.0165 & 0.947 & -0.0008 & 0.0165 & 0.0166 & 0.0165 & 0.949 \\
 & $\sigma_{b0}$ & 0.0114 & 0.3726 & 0.3699 & 0.3726 & 0.950 & 0.0162 & 0.3621 & 0.3628 & 0.3623 & 0.954 & 0.0199 & 0.3471 & 0.3580 & 0.3475 & 0.954 & 0.0156 & 0.3575 & 0.3630 & 0.3576 & 0.951 \\
 & $\sigma_{b1}$ & 0.0007 & 0.0082 & 0.0074 & 0.0082 & 0.947 & 0.0006 & 0.0078 & 0.0072 & 0.0078 & 0.948 & 0.0005 & 0.0071 & 0.0070 & 0.0071 & 0.947 & 0.0006 & 0.0077 & 0.0072 & 0.0077 & 0.945 \\
 & $\sigma_{\epsilon}$ & -0.0008 & 0.0746 & 0.0649 & 0.0746 & 0.950 & -0.0017 & 0.0688 & 0.0634 & 0.0688 & 0.950 & -0.0034 & 0.0598 & 0.0608 & 0.0598 & 0.950 & -0.0016 & 0.0705 & 0.0635 & 0.0705 & 0.952 \\
 & $\alpha$ & 0.0005 & 0.0028 & 0.0027 & 0.0028 & 0.931 & 0.0005 & 0.0028 & 0.0027 & 0.0028 & 0.933 & 0.0005 & 0.0028 & 0.0027 & 0.0028 & 0.933 & 0.0005 & 0.0028 & 0.0027 & 0.0028 & 0.933 \\
 & $\gamma$ & -0.0041 & 0.0930 & 0.0888 & 0.0931 & 0.938 & -0.0041 & 0.0932 & 0.0888 & 0.0932 & 0.941 & -0.0041 & 0.0931 & 0.0888 & 0.0932 & 0.941 & -0.0041 & 0.0932 & 0.0889 & 0.0933 & 0.939 \\
\addlinespace
2 & $\beta_0$ & -0.0073 & 0.4768 & 0.4728 & 0.4766 & 0.946 & -0.0093 & 0.4760 & 0.4728 & 0.4759 & 0.947 & -0.0095 & 0.4783 & 0.4726 & 0.4781 & 0.945 & -0.0099 & 0.4771 & 0.4726 & 0.4769 & 0.945 \\
 & $\beta_1$ & 0.0000 & 0.0121 & 0.0120 & 0.0121 & 0.950 & 0.0000 & 0.0121 & 0.0120 & 0.0121 & 0.948 & -0.0000 & 0.0121 & 0.0120 & 0.0121 & 0.952 & 0.0000 & 0.0120 & 0.0120 & 0.0120 & 0.954 \\
 & $\beta_2$ & -0.0004 & 0.0153 & 0.0156 & 0.0153 & 0.952 & -0.0004 & 0.0153 & 0.0156 & 0.0153 & 0.951 & -0.0003 & 0.0153 & 0.0156 & 0.0153 & 0.954 & -0.0004 & 0.0153 & 0.0156 & 0.0153 & 0.955 \\
 & $\sigma_{b0}$ & -0.0273 & 0.4315 & 0.3988 & 0.4321 & 0.955 & -0.0148 & 0.4088 & 0.3773 & 0.4089 & 0.951 & -0.0247 & 0.4512 & 0.3959 & 0.4517 & 0.950 & -0.0197 & 0.4225 & 0.3874 & 0.4228 & 0.953 \\
 & $\sigma_{b1}$ & 0.0011 & 0.0087 & 0.0072 & 0.0087 & 0.952 & 0.0009 & 0.0074 & 0.0067 & 0.0074 & 0.955 & 0.0011 & 0.0082 & 0.0071 & 0.0083 & 0.952 & 0.0010 & 0.0078 & 0.0070 & 0.0079 & 0.954 \\
 & $\sigma_{\epsilon}$ & 0.0072 & 0.0909 & 0.0683 & 0.0912 & 0.960 & 0.0035 & 0.0747 & 0.0622 & 0.0748 & 0.958 & 0.0067 & 0.0857 & 0.0674 & 0.0859 & 0.961 & 0.0055 & 0.0781 & 0.0658 & 0.0783 & 0.960 \\
 & $\alpha$ & 0.0004 & 0.0028 & 0.0027 & 0.0029 & 0.936 & 0.0004 & 0.0028 & 0.0027 & 0.0029 & 0.935 & 0.0004 & 0.0028 & 0.0027 & 0.0029 & 0.933 & 0.0004 & 0.0028 & 0.0027 & 0.0029 & 0.936 \\
 & $\gamma$ & -0.0075 & 0.0919 & 0.0932 & 0.0921 & 0.951 & -0.0076 & 0.0919 & 0.0932 & 0.0921 & 0.954 & -0.0075 & 0.0918 & 0.0932 & 0.0921 & 0.954 & -0.0076 & 0.0920 & 0.0933 & 0.0922 & 0.952 \\
\addlinespace
3 & $\beta_0$ & -0.0004 & 0.4668 & 0.4739 & 0.4666 & 0.953 & 0.0001 & 0.4677 & 0.4733 & 0.4674 & 0.953 & 0.0007 & 0.4684 & 0.4735 & 0.4681 & 0.950 & -0.0008 & 0.4665 & 0.4735 & 0.4663 & 0.949 \\
 & $\beta_1$ & 0.0002 & 0.0120 & 0.0120 & 0.0120 & 0.950 & 0.0003 & 0.0119 & 0.0120 & 0.0119 & 0.947 & 0.0002 & 0.0120 & 0.0120 & 0.0120 & 0.947 & 0.0002 & 0.0120 & 0.0120 & 0.0120 & 0.950 \\
 & $\beta_2$ & -0.0005 & 0.0155 & 0.0155 & 0.0155 & 0.949 & -0.0005 & 0.0154 & 0.0156 & 0.0154 & 0.952 & -0.0005 & 0.0155 & 0.0156 & 0.0155 & 0.952 & -0.0004 & 0.0155 & 0.0155 & 0.0155 & 0.946 \\
 & $\sigma_{b0}$ & 0.0083 & 0.3803 & 0.3653 & 0.3802 & 0.943 & 0.0080 & 0.3875 & 0.3650 & 0.3874 & 0.944 & -0.0162 & 0.4678 & 0.4044 & 0.4678 & 0.943 & 0.0042 & 0.3999 & 0.3716 & 0.3997 & 0.944 \\
 & $\sigma_{b1}$ & 0.0006 & 0.0073 & 0.0066 & 0.0073 & 0.945 & 0.0006 & 0.0069 & 0.0065 & 0.0070 & 0.943 & 0.0010 & 0.0083 & 0.0072 & 0.0083 & 0.945 & 0.0006 & 0.0070 & 0.0066 & 0.0070 & 0.944 \\
 & $\sigma_{\epsilon}$ & 0.0044 & 0.0700 & 0.0609 & 0.0701 & 0.943 & 0.0033 & 0.0668 & 0.0592 & 0.0668 & 0.945 & 0.0094 & 0.0920 & 0.0691 & 0.0924 & 0.946 & 0.0040 & 0.0695 & 0.0604 & 0.0696 & 0.947 \\
 & $\alpha$ & 0.0005 & 0.0030 & 0.0027 & 0.0030 & 0.926 & 0.0005 & 0.0030 & 0.0027 & 0.0030 & 0.927 & 0.0005 & 0.0030 & 0.0027 & 0.0030 & 0.927 & 0.0005 & 0.0030 & 0.0027 & 0.0030 & 0.925 \\
 & $\gamma$ & -0.0042 & 0.0938 & 0.0933 & 0.0938 & 0.939 & -0.0043 & 0.0938 & 0.0932 & 0.0938 & 0.943 & -0.0041 & 0.0938 & 0.0932 & 0.0939 & 0.950 & -0.0042 & 0.0939 & 0.0933 & 0.0940 & 0.942 \\
\addlinespace
4 & $\beta_0$ & 0.0279 & 0.4800 & 0.4798 & 0.4806 & 0.948 & 0.0278 & 0.4800 & 0.4792 & 0.4806 & 0.948 & 0.0276 & 0.4793 & 0.4794 & 0.4799 & 0.948 & 0.0282 & 0.4790 & 0.4792 & 0.4796 & 0.950 \\
 & $\beta_1$ & -0.0004 & 0.0121 & 0.0120 & 0.0121 & 0.944 & -0.0004 & 0.0121 & 0.0119 & 0.0121 & 0.945 & -0.0004 & 0.0121 & 0.0120 & 0.0121 & 0.944 & -0.0004 & 0.0121 & 0.0119 & 0.0121 & 0.946 \\
 & $\beta_2$ & -0.0001 & 0.0194 & 0.0193 & 0.0194 & 0.944 & -0.0001 & 0.0194 & 0.0193 & 0.0194 & 0.942 & -0.0001 & 0.0194 & 0.0193 & 0.0194 & 0.942 & -0.0001 & 0.0194 & 0.0193 & 0.0193 & 0.947 \\
 & $\sigma_{b0}$ & 0.0052 & 0.3735 & 0.3698 & 0.3734 & 0.965 & 0.0118 & 0.3476 & 0.3599 & 0.3476 & 0.967 & 0.0088 & 0.3566 & 0.3638 & 0.3565 & 0.967 & 0.0117 & 0.3485 & 0.3600 & 0.3486 & 0.966 \\
 & $\sigma_{b1}$ & 0.0005 & 0.0085 & 0.0082 & 0.0085 & 0.953 & 0.0003 & 0.0077 & 0.0079 & 0.0077 & 0.953 & 0.0004 & 0.0080 & 0.0080 & 0.0080 & 0.951 & 0.0003 & 0.0077 & 0.0079 & 0.0077 & 0.954 \\
 & $\sigma_{\epsilon}$ & 0.0015 & 0.0739 & 0.0689 & 0.0739 & 0.949 & -0.0000 & 0.0651 & 0.0664 & 0.0651 & 0.952 & 0.0007 & 0.0689 & 0.0676 & 0.0688 & 0.950 & -0.0000 & 0.0651 & 0.0664 & 0.0650 & 0.949 \\
 & $\alpha$ & 0.0007 & 0.0027 & 0.0027 & 0.0028 & 0.950 & 0.0007 & 0.0027 & 0.0027 & 0.0028 & 0.945 & 0.0007 & 0.0027 & 0.0027 & 0.0028 & 0.946 & 0.0007 & 0.0027 & 0.0027 & 0.0028 & 0.946 \\
 & $\gamma$ & -0.0044 & 0.0916 & 0.0889 & 0.0917 & 0.951 & -0.0044 & 0.0917 & 0.0888 & 0.0918 & 0.948 & -0.0044 & 0.0916 & 0.0887 & 0.0916 & 0.947 & -0.0044 & 0.0916 & 0.0888 & 0.0917 & 0.949 \\
\addlinespace
5 & $\beta_0$ & 0.0417 & 0.4938 & 0.4799 & 0.4953 & 0.937 & 0.0421 & 0.4936 & 0.4793 & 0.4952 & 0.938 & 0.0424 & 0.4939 & 0.4795 & 0.4955 & 0.940 & 0.0415 & 0.4933 & 0.4794 & 0.4948 & 0.940 \\
 & $\beta_1$ & 0.0002 & 0.0121 & 0.0120 & 0.0121 & 0.950 & 0.0002 & 0.0121 & 0.0120 & 0.0121 & 0.948 & 0.0002 & 0.0121 & 0.0119 & 0.0121 & 0.951 & 0.0002 & 0.0121 & 0.0120 & 0.0121 & 0.949 \\
 & $\beta_2$ & -0.0011 & 0.0197 & 0.0194 & 0.0197 & 0.949 & -0.0011 & 0.0197 & 0.0194 & 0.0197 & 0.950 & -0.0011 & 0.0197 & 0.0194 & 0.0197 & 0.946 & -0.0011 & 0.0197 & 0.0194 & 0.0197 & 0.949 \\
 & $\sigma_{b0}$ & 0.0152 & 0.3518 & 0.3600 & 0.3520 & 0.950 & 0.0117 & 0.3671 & 0.3666 & 0.3671 & 0.949 & 0.0149 & 0.3521 & 0.3597 & 0.3523 & 0.949 & 0.0155 & 0.3517 & 0.3599 & 0.3519 & 0.951 \\
 & $\sigma_{b1}$ & 0.0007 & 0.0080 & 0.0079 & 0.0080 & 0.945 & 0.0007 & 0.0084 & 0.0081 & 0.0085 & 0.947 & 0.0007 & 0.0080 & 0.0080 & 0.0080 & 0.947 & 0.0007 & 0.0080 & 0.0079 & 0.0080 & 0.948 \\
 & $\sigma_{\epsilon}$ & -0.0029 & 0.0664 & 0.0666 & 0.0665 & 0.947 & -0.0021 & 0.0709 & 0.0679 & 0.0709 & 0.954 & -0.0029 & 0.0665 & 0.0665 & 0.0665 & 0.950 & -0.0029 & 0.0664 & 0.0666 & 0.0664 & 0.954 \\
 & $\alpha$ & 0.0005 & 0.0027 & 0.0027 & 0.0028 & 0.943 & 0.0006 & 0.0027 & 0.0027 & 0.0028 & 0.945 & 0.0006 & 0.0027 & 0.0027 & 0.0028 & 0.943 & 0.0006 & 0.0027 & 0.0027 & 0.0028 & 0.948 \\
 & $\gamma$ & -0.0063 & 0.0871 & 0.0886 & 0.0873 & 0.952 & -0.0064 & 0.0871 & 0.0887 & 0.0873 & 0.956 & -0.0063 & 0.0871 & 0.0886 & 0.0872 & 0.953 & -0.0062 & 0.0872 & 0.0887 & 0.0874 & 0.955 \\
\bottomrule
\end{tabular}%
\vspace{0.5ex}
\begin{minipage}{0.95\linewidth}
\footnotesize Notes: $\text{CM}_1$ = administrative censoring. $\sigma_{\alpha}$ denotes the prior SD for $\alpha$, with $\sigma_{\alpha}=\log(F_{\max})/1.96$ for $F_{\max}=1.25,1.5,2,3$. ESD = empirical SD; PSD = posterior SD; CP = coverage probability. Values summarize the 1000 simulation replicates within each log-logistic data-generation scenario.
\end{minipage}
\end{table}

\end{landscape}

\begin{landscape}
\begin{table}[!htbp]
\centering
\tiny
\caption{Detailed $\alpha$-prior sensitivity statistics for log-logistic generated data under controlled 50\% censoring.}
\label{tab:alpha-prior-sensitivity-details-cm2}
\begin{tabular}{llrrrrrrrrrrrrrrrrrrrr}
\toprule
Scen. & Par. & \multicolumn{5}{c}{$\sigma_{\alpha}=0.114$} & \multicolumn{5}{c}{$\sigma_{\alpha}=0.207$} & \multicolumn{5}{c}{$\sigma_{\alpha}=0.354$} & \multicolumn{5}{c}{$\sigma_{\alpha}=0.561$}
\\
\cmidrule(lr){3-7} \cmidrule(lr){8-12} \cmidrule(lr){13-17} \cmidrule(lr){18-22}
 &  & Bias & ESD & PSD & RMSE & CP & Bias & ESD & PSD & RMSE & CP & Bias & ESD & PSD & RMSE & CP & Bias & ESD & PSD & RMSE & CP
\\
\midrule
1 & $\beta_0$ & 0.0073 & 0.4788 & 0.4803 & 0.4786 & 0.948 & 0.0084 & 0.4780 & 0.4803 & 0.4778 & 0.945 & 0.0067 & 0.4795 & 0.4804 & 0.4793 & 0.949 & 0.0080 & 0.4785 & 0.4808 & 0.4784 & 0.945 \\
 & $\beta_1$ & -0.0005 & 0.0146 & 0.0144 & 0.0146 & 0.949 & -0.0005 & 0.0147 & 0.0144 & 0.0147 & 0.948 & -0.0004 & 0.0146 & 0.0144 & 0.0146 & 0.949 & -0.0004 & 0.0146 & 0.0144 & 0.0146 & 0.949 \\
 & $\beta_2$ & 0.0005 & 0.0200 & 0.0200 & 0.0200 & 0.938 & 0.0005 & 0.0200 & 0.0201 & 0.0200 & 0.939 & 0.0005 & 0.0200 & 0.0200 & 0.0199 & 0.941 & 0.0005 & 0.0200 & 0.0201 & 0.0200 & 0.943 \\
 & $\sigma_{b0}$ & -0.0120 & 0.3717 & 0.3622 & 0.3717 & 0.940 & -0.0152 & 0.3797 & 0.3680 & 0.3799 & 0.941 & -0.0116 & 0.3719 & 0.3620 & 0.3719 & 0.941 & -0.0123 & 0.3715 & 0.3621 & 0.3715 & 0.946 \\
 & $\sigma_{b1}$ & -0.0001 & 0.0091 & 0.0090 & 0.0091 & 0.952 & 0.0001 & 0.0099 & 0.0093 & 0.0099 & 0.951 & -0.0001 & 0.0091 & 0.0090 & 0.0091 & 0.951 & -0.0001 & 0.0091 & 0.0090 & 0.0091 & 0.950 \\
 & $\sigma_{\epsilon}$ & 0.0035 & 0.0701 & 0.0714 & 0.0702 & 0.955 & 0.0051 & 0.0783 & 0.0738 & 0.0784 & 0.956 & 0.0036 & 0.0701 & 0.0714 & 0.0701 & 0.952 & 0.0036 & 0.0700 & 0.0713 & 0.0701 & 0.954 \\
 & $\alpha$ & 0.0007 & 0.0033 & 0.0032 & 0.0034 & 0.940 & 0.0007 & 0.0033 & 0.0032 & 0.0034 & 0.944 & 0.0007 & 0.0033 & 0.0032 & 0.0034 & 0.944 & 0.0007 & 0.0033 & 0.0032 & 0.0034 & 0.943 \\
 & $\gamma$ & -0.0035 & 0.1012 & 0.1001 & 0.1012 & 0.950 & -0.0035 & 0.1013 & 0.1001 & 0.1013 & 0.948 & -0.0035 & 0.1013 & 0.1001 & 0.1013 & 0.946 & -0.0036 & 0.1013 & 0.1002 & 0.1013 & 0.950 \\
\addlinespace
2 & $\beta_0$ & 0.0187 & 0.4841 & 0.4756 & 0.4842 & 0.947 & 0.0194 & 0.4841 & 0.4750 & 0.4843 & 0.950 & 0.0183 & 0.4840 & 0.4755 & 0.4841 & 0.950 & 0.0194 & 0.4831 & 0.4753 & 0.4833 & 0.943 \\
 & $\beta_1$ & 0.0001 & 0.0127 & 0.0130 & 0.0127 & 0.959 & 0.0002 & 0.0127 & 0.0129 & 0.0127 & 0.959 & 0.0001 & 0.0126 & 0.0129 & 0.0126 & 0.959 & 0.0001 & 0.0127 & 0.0129 & 0.0127 & 0.960 \\
 & $\beta_2$ & -0.0001 & 0.0169 & 0.0168 & 0.0169 & 0.950 & -0.0002 & 0.0169 & 0.0168 & 0.0169 & 0.952 & -0.0001 & 0.0168 & 0.0168 & 0.0168 & 0.953 & -0.0001 & 0.0169 & 0.0168 & 0.0169 & 0.950 \\
 & $\sigma_{b0}$ & -0.0114 & 0.3864 & 0.3708 & 0.3864 & 0.951 & -0.0079 & 0.3769 & 0.3653 & 0.3768 & 0.956 & -0.0118 & 0.3964 & 0.3722 & 0.3963 & 0.948 & -0.0039 & 0.3535 & 0.3582 & 0.3534 & 0.955 \\
 & $\sigma_{b1}$ & 0.0011 & 0.0079 & 0.0073 & 0.0080 & 0.946 & 0.0011 & 0.0077 & 0.0072 & 0.0078 & 0.952 & 0.0012 & 0.0082 & 0.0073 & 0.0083 & 0.950 & 0.0010 & 0.0075 & 0.0071 & 0.0076 & 0.950 \\
 & $\sigma_{\epsilon}$ & 0.0007 & 0.0737 & 0.0647 & 0.0736 & 0.948 & -0.0001 & 0.0703 & 0.0634 & 0.0702 & 0.948 & 0.0007 & 0.0726 & 0.0647 & 0.0725 & 0.947 & -0.0008 & 0.0662 & 0.0623 & 0.0662 & 0.947 \\
 & $\alpha$ & 0.0006 & 0.0031 & 0.0030 & 0.0032 & 0.935 & 0.0006 & 0.0031 & 0.0030 & 0.0032 & 0.938 & 0.0006 & 0.0031 & 0.0030 & 0.0032 & 0.936 & 0.0006 & 0.0031 & 0.0030 & 0.0032 & 0.932 \\
 & $\gamma$ & -0.0023 & 0.1002 & 0.1000 & 0.1002 & 0.945 & -0.0022 & 0.1001 & 0.1000 & 0.1001 & 0.944 & -0.0023 & 0.1001 & 0.1001 & 0.1001 & 0.949 & -0.0022 & 0.1001 & 0.1001 & 0.1001 & 0.947 \\
\addlinespace
3 & $\beta_0$ & -0.0095 & 0.4810 & 0.4765 & 0.4809 & 0.948 & -0.0108 & 0.4822 & 0.4762 & 0.4821 & 0.944 & -0.0106 & 0.4812 & 0.4763 & 0.4811 & 0.942 & -0.0108 & 0.4822 & 0.4765 & 0.4821 & 0.947 \\
 & $\beta_1$ & -0.0005 & 0.0131 & 0.0130 & 0.0131 & 0.954 & -0.0005 & 0.0131 & 0.0130 & 0.0131 & 0.949 & -0.0005 & 0.0131 & 0.0130 & 0.0131 & 0.946 & -0.0005 & 0.0131 & 0.0130 & 0.0131 & 0.947 \\
 & $\beta_2$ & -0.0001 & 0.0169 & 0.0169 & 0.0169 & 0.949 & -0.0001 & 0.0170 & 0.0170 & 0.0170 & 0.952 & -0.0000 & 0.0169 & 0.0169 & 0.0168 & 0.951 & -0.0000 & 0.0169 & 0.0169 & 0.0169 & 0.952 \\
 & $\sigma_{b0}$ & 0.0050 & 0.4187 & 0.3805 & 0.4185 & 0.942 & 0.0024 & 0.4277 & 0.3841 & 0.4275 & 0.938 & 0.0138 & 0.3846 & 0.3669 & 0.3847 & 0.938 & 0.0010 & 0.4356 & 0.3870 & 0.4354 & 0.941 \\
 & $\sigma_{b1}$ & 0.0006 & 0.0087 & 0.0077 & 0.0087 & 0.945 & 0.0007 & 0.0090 & 0.0077 & 0.0090 & 0.946 & 0.0004 & 0.0079 & 0.0073 & 0.0079 & 0.946 & 0.0007 & 0.0091 & 0.0078 & 0.0091 & 0.947 \\
 & $\sigma_{\epsilon}$ & 0.0050 & 0.0790 & 0.0674 & 0.0791 & 0.957 & 0.0055 & 0.0813 & 0.0685 & 0.0815 & 0.957 & 0.0029 & 0.0684 & 0.0641 & 0.0685 & 0.959 & 0.0057 & 0.0811 & 0.0687 & 0.0813 & 0.959 \\
 & $\alpha$ & 0.0006 & 0.0032 & 0.0030 & 0.0033 & 0.926 & 0.0006 & 0.0032 & 0.0030 & 0.0033 & 0.924 & 0.0006 & 0.0032 & 0.0030 & 0.0033 & 0.926 & 0.0006 & 0.0032 & 0.0030 & 0.0033 & 0.927 \\
 & $\gamma$ & -0.0021 & 0.1002 & 0.1003 & 0.1001 & 0.957 & -0.0021 & 0.1003 & 0.1003 & 0.1003 & 0.957 & -0.0020 & 0.1003 & 0.1003 & 0.1003 & 0.957 & -0.0020 & 0.1002 & 0.1004 & 0.1002 & 0.961 \\
\addlinespace
4 & $\beta_0$ & 0.0241 & 0.4951 & 0.4870 & 0.4954 & 0.949 & 0.0231 & 0.4946 & 0.4869 & 0.4949 & 0.948 & 0.0241 & 0.4952 & 0.4879 & 0.4955 & 0.942 & 0.0235 & 0.4955 & 0.4871 & 0.4958 & 0.946 \\
 & $\beta_1$ & -0.0013 & 0.0164 & 0.0164 & 0.0165 & 0.947 & -0.0013 & 0.0164 & 0.0164 & 0.0164 & 0.948 & -0.0014 & 0.0164 & 0.0164 & 0.0164 & 0.949 & -0.0013 & 0.0164 & 0.0164 & 0.0164 & 0.948 \\
 & $\beta_2$ & -0.0008 & 0.0255 & 0.0260 & 0.0255 & 0.958 & -0.0007 & 0.0254 & 0.0260 & 0.0254 & 0.957 & -0.0007 & 0.0254 & 0.0260 & 0.0254 & 0.956 & -0.0008 & 0.0254 & 0.0260 & 0.0254 & 0.958 \\
 & $\sigma_{b0}$ & 0.0083 & 0.3691 & 0.3720 & 0.3691 & 0.947 & 0.0080 & 0.3684 & 0.3718 & 0.3683 & 0.948 & 0.0076 & 0.3693 & 0.3718 & 0.3692 & 0.947 & 0.0077 & 0.3688 & 0.3719 & 0.3687 & 0.947 \\
 & $\sigma_{b1}$ & 0.0006 & 0.0121 & 0.0122 & 0.0121 & 0.954 & 0.0006 & 0.0121 & 0.0122 & 0.0121 & 0.954 & 0.0006 & 0.0121 & 0.0122 & 0.0121 & 0.955 & 0.0006 & 0.0121 & 0.0122 & 0.0121 & 0.953 \\
 & $\sigma_{\epsilon}$ & -0.0018 & 0.0835 & 0.0851 & 0.0835 & 0.956 & -0.0018 & 0.0835 & 0.0852 & 0.0835 & 0.957 & -0.0019 & 0.0835 & 0.0852 & 0.0835 & 0.958 & -0.0019 & 0.0835 & 0.0852 & 0.0835 & 0.956 \\
 & $\alpha$ & 0.0009 & 0.0035 & 0.0035 & 0.0036 & 0.941 & 0.0009 & 0.0035 & 0.0035 & 0.0036 & 0.940 & 0.0009 & 0.0035 & 0.0035 & 0.0036 & 0.937 & 0.0009 & 0.0035 & 0.0035 & 0.0036 & 0.943 \\
 & $\gamma$ & -0.0079 & 0.1037 & 0.1037 & 0.1039 & 0.943 & -0.0080 & 0.1037 & 0.1037 & 0.1040 & 0.943 & -0.0080 & 0.1037 & 0.1037 & 0.1040 & 0.941 & -0.0081 & 0.1037 & 0.1037 & 0.1039 & 0.948 \\
\addlinespace
5 & $\beta_0$ & 0.0065 & 0.4903 & 0.4864 & 0.4901 & 0.947 & 0.0059 & 0.4889 & 0.4867 & 0.4886 & 0.948 & 0.0055 & 0.4898 & 0.4866 & 0.4895 & 0.948 & 0.0043 & 0.4899 & 0.4865 & 0.4896 & 0.946 \\
 & $\beta_1$ & -0.0020 & 0.0169 & 0.0164 & 0.0170 & 0.933 & -0.0021 & 0.0170 & 0.0164 & 0.0171 & 0.936 & -0.0021 & 0.0169 & 0.0164 & 0.0170 & 0.933 & -0.0020 & 0.0169 & 0.0164 & 0.0170 & 0.933 \\
 & $\beta_2$ & 0.0007 & 0.0269 & 0.0262 & 0.0269 & 0.949 & 0.0007 & 0.0269 & 0.0262 & 0.0269 & 0.947 & 0.0007 & 0.0269 & 0.0262 & 0.0269 & 0.949 & 0.0007 & 0.0269 & 0.0262 & 0.0269 & 0.947 \\
 & $\sigma_{b0}$ & -0.0097 & 0.3792 & 0.3717 & 0.3791 & 0.941 & -0.0089 & 0.3798 & 0.3713 & 0.3797 & 0.941 & -0.0089 & 0.3798 & 0.3716 & 0.3797 & 0.944 & -0.0099 & 0.3791 & 0.3717 & 0.3790 & 0.942 \\
 & $\sigma_{b1}$ & 0.0006 & 0.0120 & 0.0122 & 0.0120 & 0.948 & 0.0007 & 0.0120 & 0.0122 & 0.0120 & 0.953 & 0.0007 & 0.0120 & 0.0122 & 0.0120 & 0.958 & 0.0007 & 0.0120 & 0.0122 & 0.0120 & 0.948 \\
 & $\sigma_{\epsilon}$ & -0.0017 & 0.0827 & 0.0855 & 0.0827 & 0.950 & -0.0017 & 0.0827 & 0.0854 & 0.0827 & 0.950 & -0.0017 & 0.0827 & 0.0854 & 0.0827 & 0.949 & -0.0017 & 0.0827 & 0.0855 & 0.0827 & 0.951 \\
 & $\alpha$ & 0.0009 & 0.0037 & 0.0035 & 0.0038 & 0.936 & 0.0009 & 0.0037 & 0.0035 & 0.0038 & 0.933 & 0.0009 & 0.0037 & 0.0035 & 0.0038 & 0.935 & 0.0009 & 0.0037 & 0.0035 & 0.0038 & 0.939 \\
 & $\gamma$ & -0.0114 & 0.1011 & 0.1036 & 0.1017 & 0.958 & -0.0115 & 0.1010 & 0.1036 & 0.1016 & 0.961 & -0.0115 & 0.1010 & 0.1036 & 0.1016 & 0.957 & -0.0114 & 0.1011 & 0.1036 & 0.1017 & 0.958 \\
\bottomrule
\end{tabular}%
\vspace{0.5ex}
\begin{minipage}{0.95\linewidth}
\footnotesize Notes: $\text{CM}_2$ = controlled 50\% censoring. $\sigma_{\alpha}$ denotes the prior SD for $\alpha$, with $\sigma_{\alpha}=\log(F_{\max})/1.96$ for $F_{\max}=1.25,1.5,2,3$. ESD = empirical SD; PSD = posterior SD; CP = coverage probability. Values summarize the 1000 simulation replicates within each log-logistic data-generation scenario.
\end{minipage}
\end{table}

\end{landscape}

\bibliographystyle{apalike}
\bibliography{references}